%% file: paper_Qtorsion.tex
\documentclass[aps,prd,reprint,preprintnumbers,showpacs,floatfix,nofootinbib,superscriptaddress]{revtex4-2}
\usepackage[utf8]{inputenc}
\usepackage{parskip}

\usepackage{amssymb}
\usepackage{amsmath}
\usepackage{mathtools}
\usepackage[dvipsnames]{xcolor}
\usepackage{xspace}
\usepackage{tabularx}
\usepackage{siunitx}
\usepackage{multirow}
\usepackage{subcaption}
\usepackage{graphicx}
\usepackage{xstring}
\usepackage{etoolbox}
\usepackage{notoccite}
\usepackage{natbib}
\usepackage{mathrsfs}
\usepackage{lineno}
\usepackage{tensor}
\usepackage{accents}
\usepackage{microtype}
\usepackage[justification=raggedright,singlelinecheck=false,format=plain]{caption}

\usepackage{tocbasic}
\DeclareTOCStyleEntry[numwidth=25pt,linefill=\bfseries\TOCLineLeaderFill]{tocline}{section}
\DeclareTOCStyleEntry[entryformat=\textit,numwidth=10pt,linefill=\TOCLineLeaderFill]{tocline}{subsection}
\DeclareTOCStyleEntry[entryformat=\textit,numwidth=10pt,linefill=\TOCLineLeaderFill]{tocline}{subsubsection}


\usepackage{hyperref}
\hypersetup{%
 colorlinks = true,%
 linkcolor = red,%
 citecolor = red,%
 filecolor = red,%
 urlcolor = red% 
 }%
\usepackage[capitalize]{cleveref}
\AddToHook{cmd/appendix/before}{\crefalias{section}{appendix}} 
\makeatletter
\renewcommand{\paragraph}{%
\@startsection{paragraph}{4}%
{\z@}{1.21ex \@plus 1ex \@minus .2ex}{0.9em}%
{\normalfont\normalsize\bfseries}%
}
\usepackage{titlesec}
\newrobustcmd{\pea}[1]{%
	\emph{#1}\textbf{\ \ \ ---}
}
\titleformat{\paragraph}[runin]{\normalfont\normalsize\bfseries}{\emph\theparagraph}{1em}{\pea}
\titleformat{\section}[block]{\normalfont\bfseries\centering}{\MakeUppercase\thesection}{1em}{\MakeUppercase}
\makeatother

\newcommand{\hunit}{km\,s$^{-1}$\,Mpc$^{-1}\,~$}
\newcommand{\LCDM}{\ensuremath{\Lambda\mathrm{CDM}}}
\newcommand{\Mp}{M_\mathrm{P}^2}
\newcommand{\RR}{\mathcal{R}}
\newcommand{\TT}{\mathcal{T}}

\newcommand{\LTorC}{\mathcal{L}_{\mathrm{TorC}}^{\mathrm{ST}}}
\DeclareSIUnit\Mpc{Mpc}

\newrobustcmd{\SmashAcute}[1]{%
	{\vphantom{#1}\smash{\Acute{#1}}}
}
\newrobustcmd{\ECT}[1]{%
	\tensor{\mathcal{T}}{#1}
}
\newrobustcmd{\ECR}[1]{%
	\tensor{\mathcal{R}}{#1}
}
\newrobustcmd{\MetricPerturbation}[1]{%
	\tensor*{h}{#1}
}
\newrobustcmd{\RiemannianR}[1]{%
	\tensor{R}{#1}
}
\newrobustcmd{\StressEnergyTensor}[1]{%
	\tensor{\mathbb{T}}{#1}
}
\newrobustcmd{\TorsionSource}[1]{%
	\tensor{\Delta}{#1}
}
\newrobustcmd{\AffineConnection}[1]{%
	\tensor{\Gamma}{#1}
}
\newrobustcmd{\LeviCivitaConnection}[1]{%
	\big\{\tensor{}{#1}\big\}
}
\newrobustcmd{\PD}[1]{%
	\tensor{\partial}{#1}
}
\newrobustcmd{\CD}[1]{%
	\tensor{\nabla}{#1}
}
\makeatletter

\allowdisplaybreaks

\begin{document}

\preprint{Published in \href{https://doi.org/10.1088/1475-7516/2026/03/003}{JCAP03(2026)003}}

\title{Alleviating the Hubble tension with Torsion Condensation (TorC)}

\author{Sinah Legner}
\email{sl2091@cam.ac.uk}
\affiliation{Astrophysics Group, Cavendish Laboratory, JJ Thomson Avenue, Cambridge CB3 0HE, UK}
\affiliation{Kavli Institute for Cosmology, Madingley Road, Cambridge CB3 0HA, UK}

\author{Will Handley}
\email{wh260@cam.ac.uk}
\affiliation{Kavli Institute for Cosmology, Madingley Road, Cambridge CB3 0HA, UK}
\affiliation{Institute of Astronomy, Madingley Road, Cambridge CB3 0HA, UK}

\author{Will Barker}
\email{barker@fzu.cz}
\affiliation{Central European Institute for Cosmolgy and Fundamental Physics, Institute of Physics of the Czech Academy of Sciences, Na Slovance 1999/2, 182 00 Prague 8, Czechia}

\author{Adam Ormondroyd}
\email{ano23@cam.ac.uk}
\affiliation{Astrophysics Group, Cavendish Laboratory, JJ Thomson Avenue, Cambridge CB3 0HE, UK}
\affiliation{Kavli Institute for Cosmology, Madingley Road, Cambridge CB3 0HA, UK}
\begin{abstract}
		Constraints on the cosmological parameters of Torsion Condensation (TorC) are investigated using Planck 2018 Cosmic Microwave Background data. TorC is a case of Poincar\'e gauge theory --- a formulation of gravity motivated by the gauge field theories underlying fundamental forces in the standard model of particle physics.  Unlike general relativity, TorC incorporates intrinsic torsion degrees of freedom while maintaining second-order field equations.  At specific parameter values, it reduces to the \LCDM{} model, providing a natural extension to standard cosmology. The base model of TorC introduces two parameters beyond those in \LCDM{}: the initial value of the torsion scalar field and its time derivative --- one can absorb the latter by allowing the dark energy density to float. To constrain these parameters, the \texttt{PolyChord} nested sampling algorithm is employed, interfaced via \texttt{Cobaya} with a modified version of \texttt{CAMB}.  Our results indicate that TorC allows for a larger inferred Hubble constant, offering a potential resolution to the Hubble tension.  Tension analysis using the~$R$-statistic shows that TorC alleviates the statistical tension between the Planck 2018 and SH0ES 2020 datasets, though this improvement is not sufficient to decisively favour TorC over \LCDM{} in a Bayesian model comparison. This study highlights TorC as a compelling theory of gravity, demonstrating its potential to address cosmological tensions and motivating further investigations of extended theories of gravity within a cosmological context. As current and upcoming surveys --- including Euclid, Roman Space Telescope, Vera C. Rubin Observatory, LISA, and Simons Observatory --- deliver data on gravity across all scales, they will offer critical tests of gravity models like TorC, making the present a pivotal moment for exploring extended theories of gravity.

\end{abstract}

\maketitle

% Reduce spacing in table of contents
{
\setlength{\parskip}{0pt}
\setlength{\topsep}{0pt}
\setlength{\partopsep}{0pt}
\setlength{\itemsep}{0pt}
\renewcommand{\baselinestretch}{0.8}
\tableofcontents
}
\setlength{\parskip}{6pt}

\section{Introduction}\label{sec:intro}

The current cosmological-constant/cold-dark-matter model of cosmology, \LCDM{}, is built upon two foundations: the theory of general relativity (GR), and the observation that the Universe becomes homogeneous and isotropic at a scale larger than approximately~\SI{100}{\Mpc}~\cite{Sarkar_2009, Scrimgeour_2012}. The model has proven success in explaining a wide range of observed phenomena, including the accelerated expansion of the Universe, anisotropy in the cosmic microwave background (CMB)~\cite{WMAP:2003tof}, and the observed abundances of deuterium, helium, and other atomic nuclei~\cite{Cyburt:2015mya}. Despite its success, the \LCDM{} model is currently under scrutiny due to the indications from observational astrophysical data that suggest potential inconsistencies. These include the Hubble tension~\cite{Riess:2019cxk, Verde:2019ivm, DiValentino:2021izs, Dainotti:2021pqg, Dainotti:2022bzg, Dainotti:2025qxz}, curvature tension~\cite{Handley:2019tkm, DiValentino:2019qzk}, and discrepancies in the light element abundances~\cite{Fields:2019pfx}. Furthermore, the theoretical provenance of dark energy in the form of a very small cosmological constant~$\Lambda$, dark matter, and the inflaton field continue to pose a mystery despite decades of research. Another challenge lies in successfully uniting the gravitational force with the other forces of nature described in the standard model (SM) of particle physics. As gravity is non-renormalisable in the framework of quantum field theory, it can unfortunately only be investigated as an effective field theory valid up to the Planck scale~\cite{Donoghue:1994dn}.\footnote{This is genuinely limiting, because the measured Higgs mass implies runnings such that the SM remains predictive up to this scale.} Together, these theoretical challenges and observational tensions motivate the exploration of modified or extended theories of gravity that may offer solutions to these problems~\cite{DeSimone:2024lvy}.

\begin{figure}[t]
\includegraphics[width=\linewidth]{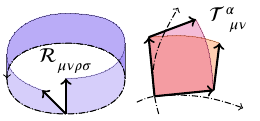}
\caption{A schematic illustration of curvature~$\tensor{\RR}{_{\mu\nu\rho\sigma}}$ and torsion~$\tensor{\TT}{^{\alpha}_{\mu\nu}}$. The left of the diagram demonstrates curvature as taught in most introductions to GR: In the presence of curvature, the direction of vector changes when parallely transported around a loop. The right of the diagram illustrates torsion: In the presence of torsion, the infinitesimal parallelogram spanned by two vectors does not close.}
\label{fig:NonRiemannianSchematic}
\end{figure}

\paragraph*{Poincar\'e gauge theory} The gauge approach to gravity stands out among various theories of gravity, as gauge theories have been very successful in explaining the electroweak and strong sectors in particle physics, where forces are mediated by vector gauge bosons. Due to the importance of global Poincar\'e symmetry --- which includes four spacetime translations and the Lorentz group containing three spatial rotations and Lorentz boosts which rotate spacetime in three spatial directions --- the consideration of Poincaré gauge theory (PGT) arises naturally as a framework for describing the gravitational force~\cite{Utiyama:1956sy, Kibble:1961ba, Sciama:1964jqa, Hehl:1976kj}. The gauge fields in PGT gauge the translational and rotational parts of the Poincaré group~\cite{Blagojevic:2002du}. The Lagrangian of PGT is built from the field strength tensors, one for each gauge field, which are identified with the torsion tensor~$\TT$, and the curvature tensor~$\RR$, respectively. The theory up to quadratic order in the curvature and torsion tensors with even parity has Lagrangian density~$\mathcal{L} \sim \RR + \RR^2 + \TT^2 + \mathcal{L}_\mathrm{M}$, where~$\mathcal{L}_\mathrm{M}$ is the matter Lagrangian. Whilst in GR,~$\RR^2$ operators are expected as one-loop corrections to the Einstein--Hilbert term in the presence of matter, in this paper, the quadratic Lagrangian structure is motivated by analogy to Yang--Mills theory since the same structures arise in the gauge theories of the SM.
\paragraph*{Can Poincar\'e gauge theory be saved?} The specific model considered in this paper is purely of the form~$\RR^2 + \TT^2$, and it has been shown to be renormalisable by power counting~\cite{Lin:2018awc, Lin:2019ugq}.\footnote{As such, the full particle content of the theory is taken as being physical: new particles in~$\RR + \RR^2 +\TT^2$ theories which have been motivated as non-renormalisable effective field theories would instead be interpreted as truncation artefacts~\cite{Glavan:2024cfs}.} While the model’s power-counting renormalizability is a promising feature, a complete demonstration of its perturbative viability requires further investigation. The analysis in~\cite{Lin:2019ugq} is based on matching canonical dimensions of modes in the linearized theory, which is a necessary but insufficient condition for renormalizability without explicit loop-level calculations. The construction of a fully consistent quantum theory of torsion, guided by Wilsonian effective field theory principles, is an active area of research~\cite{Barker:2025xzd,Barker:2025rzd,Barker:2025fgo} (see also groundwork in~\cite{Baldazzi:2021kaf,Martini:2023apm,Sauro:2022chz,Sauro:2024ujx,Melichev:2023lwj,Melichev:2025hcg}). As shown in~\cref{sec:perturbations}, the free limit of the theory considered exhibits many symmetries. However, they appear to arise simply because higher-rank and higher-spin components of the torsion tensor do not participate in the simplified quadratic approximation, as indicated by the theory's small and unitary spectrum.  The crucial problem is that such accidental symmetries are likely broken in the full non-linear completion proposed by~\cite{Lin:2019ugq}. The breaking of accidental symmetries is a signal of strong coupling, which is fatal to perturbation theory. A consistent effective field theory, as in~\cite{Barker:2025fgo}, would preserve the fundamental symmetries to avoid such pathologies.

\paragraph*{Farewell to perturbativity} 
While the above considerations challenge the perturbation theory for general theories with torsion~\cite{Barker:2025xzd}, the particular model from~\cite{Lin:2019ugq} studied in this paper is not easily dismissed. Returning to tree level, it is explained heuristically in~\cref{sec:perturbations} --- and shown in separate work~\cite{Barker:2025} --- that the theory is expected to retain an Einsteinian limit even in the absence of the Einstein--Hilbert term. This is due to the post-Riemannian expansion of the Einstein--Cartan curvature, which gives rise to a cross-term~$\RR \TT^2$. Condensation of the torsion field at~$\TT^2 \sim \Mp$ leads to the emergence of the Einstein--Hilbert term. This condensation is an inherently non-perturbative feature. Consequently, if the theory is physically realized only in this ``condensed phase'', the properties of perturbation theory around the na\"ive zero-torsion background become less critical. For instance, while the perturbative theory is unitary, this is not necessary for the model's viability. A useful parallel is the Higgs sector in the Standard Model, which appears tachyonic before the electroweak transition but is unitary in its true, symmetry-broken ground state. In fact, the very pathologies conjectured earlier, such as strongly-coupled modes, could be what makes the theory's behavior near~$\TT\sim 0$ inherently non-perturbative. Such features have precedent in theories like $f(R)$ gravity at~$f'(R)=0$~\cite{Barker:2025gon}, where so long as these surfaces are not phenomenologically relevant, the theory can remain healthy.

\paragraph*{Torsion condensation (TorC)} Let us now discuss the model of this paper in more specific detail. The method of spin projection operators was used to study PGT systematically, identifying parameter conditions that avoid unphysical ghost and tachyon degrees of freedom~\cite{Lin:2018awc, Lin:2019ugq}. The most developed among the resulting catalogue of models is the Torsion Condensation (TorC) model~\cite{Barker:2020elg, Barker:2020gcp, barker_2021, Rew:2023zxy}, with the Lagrangian density
\begin{equation}\label{eq:TorCLagrangian}
\begin{aligned}
\mathcal{L}_{\mathrm{TorC}} = & -\frac{4\Mp}{9} \tensor{\TT}{_\mu} \tensor{\TT}{^\mu} \\
& - \frac{\mu}{6} \Big[ \lambda \tensor{\TT}{_{\mu\nu\sigma}}\big( \tensor{\TT}{^{\mu\nu\sigma}} - 2 \tensor{\TT}{^{\nu\mu\sigma}} \big) \\
& + 12\tensor{\RR}{_{\mu\nu}} \big( \tensor{\RR}{^{[\mu\nu]}} - \tensor{\RR}{^{\mu\nu}} \big) \\
& - 2 \tensor{\RR}{_{\mu\nu\sigma\rho}} \big( \tensor{\RR}{^{\mu\nu\sigma\rho}} - 4 \tensor{\RR}{^{\mu\sigma\nu\rho}} - 5 \tensor{\RR}{^{\sigma\rho\mu\nu}} \big) \Big] \\
& + 2 \nu \tensor{\RR}{_{[\mu\nu]}} \tensor{\RR}{^{[\mu\nu]}} - \Mp \Lambda + \mathcal{L}_{\mathrm{M}},
\end{aligned}
\end{equation}
where~$\tensor{\RR}{_{\mu\nu\sigma\rho}}$ is the curvature, and~$\tensor{\TT}{^\mu_{\nu\sigma}}$ is the torsion tensor, with~$\tensor{\TT}{_\nu} \equiv \tensor{\TT}{^{\mu}_{\nu\mu}}$;~$\Mp = 1/8\pi G$ is the Planck mass,~$\mu$ and~$\nu$ are dimensionless curvature coupling constants;~$\lambda$ is a coupling constant of mass dimension two, which can be shown to give rise to a torsional dark energy term intrinsic to the gravitational sector, whereas~$\Lambda$ is the usual cosmological constant (which could also be absorbed into the matter Lagrangian~$\mathcal{L}_{\mathrm{M}}$). As shown in~\cref{sec:perturbations}, to ensure the absence of ghost and tachyon degrees of freedom in the tree-level perturbation theory, the following inequality must be satisfied~\cite{Lin:2019ugq}
\begin{equation}
	\lambda \geq 0, \quad \mu < 0, \quad (\nu + 2 \mu)(\nu - \mu) > 0.
\label{eq:TorCconditions}
\end{equation}
As discussed above, the conditions in~\cref{eq:TorCconditions} were originally derived with reference to the quadratic theory of perturbations around the~$\TT=0$ background: the corresponding conditions for the condensate are not yet known. Only the parameters~$\lambda$ and~$\mu$, are relevant to condensation of the background cosmology, and we will see presently that the constraints in~\cref{eq:TorCconditions} are perfectly compatible with this process. In particular, we assume the constraint
\begin{equation}
	\lambda = 0,
\end{equation}
which eliminates the torsional dark energy~\cite{Barker:2020elg}, reducing the dark energy sector to the standard cosmological constant,~$\Lambda$. Future work will explore the alternative scenario where~$\lambda$ itself drives cosmic acceleration, potentially eliminating the need for a cosmological constant.

An FLRW background is assumed, characterized by the line element
\begin{equation}
	\mathrm{d}s^2 = -\mathrm{d}t^2 + a^2(t) \left(\frac{\mathrm{d}r^2}{1 - kr^2} + r^2 \mathrm{d}\Omega^2\right),
\end{equation}
where~$k$ is the dimensionful curvature parameter, and~$a(t)$ is the dimensionless scale factor of the Universe. It can be shown that the TorC field equations are insensitive to the value of~$k$, so that the expansion history is not affected by the actual closed, flat or open spatial geometry of the Universe.\footnote{Note that the full implications of this property --- which was termed `$k$-screening' in~\cite{Barker:2020gcp} --- are not yet understood. Insensitivity to $k$ is an exact symmetry of the fully non-linear background (minisuperspace) equations.} Therefore, a flat FLRW background is adopted throughout this work.

\paragraph*{Scalar-tensor equivalent of TorC} To explore the cosmology of TorC, an equivalent scalar-tensor formulation is introduced. Among the various gravity theories, those based on scalar fields within the framework of modified gravity are one of the most extensively explored models. Scalar-tensor theories involve the coupling of different scalar fields to the metric in a curved but torsion-free spacetime~\cite{Horndeski:1974wa}. Note that, in the absence of torsion, the non-Riemannian curvature~$\ECR{}$ becomes the usual Riemannian curvature~$\RiemannianR{}$ familiar from GR --- this matter is explained in detail in~\cref{sec:perturbations}. In the cosmological context, scalar fields offer the advantage of inducing accelerated expansion without necessarily violating the isotropy of the Universe,\footnote{This is in contrast to vector models, which always `point somewhere'; the condensation of a vector field to finite vacuum expectation value is generally understood to constitute an \ae{}ther~\cite{Skordis:2020eui, Hsu:2024ftc}.} making them compelling candidates for dark energy and early-Universe inflationary mechanisms. It has been shown that in a flat, homogeneous and isotropic Universe, the Lagrangian of PGT can be mapped to a bi-scalar-tensor theory~\cite{Barker:2020elg}. For TorC, the scalar-tensor equivalent Lagrangian density is
\begin{equation}\label{TorCLagrangianMA}
\begin{aligned}
	\LTorC &= \Mp \Biggl[-\frac{2}{3} \Bigl(1-\frac{\varpi^2}{4}\Bigr) \RiemannianR{} + g^{\mu\nu} \partial_{\mu} \varpi \partial_{\nu} \varpi \\
		&\ -\frac{4}{3} \sqrt{|J_{\mu } J^{\mu }|} - \phi^2 + \phi^2 \varpi^2 \Biggr] - \Mp \Lambda + \mathcal{L}_\mathrm{M}, \\
		J_{\mu} &\equiv 4 \varpi^3 \partial_{\mu} (\phi/\varpi) + \partial_{\mu} \phi .
\end{aligned}
\end{equation}
In~\cref{TorCLagrangianMA}, both~$\varpi$ and~$\phi$ are scalar fields that arise naturally from the unique form adopted by torsion in a homogeneous and isotropic Universe~\cite{TSAMPARLIS197927},\footnote{Whilst~\cref{Tsamparlis} looks complicated, the principle for deriving it is very simple. The torsion tensor has index symmetry~$\TT{^\mu_{\nu\sigma}}=-\TT{^\mu_{\sigma\nu}}$. At the level of the background cosmology, the expanding Universe provides a preferred frame such that the index `0' corresponds to time experienced by observers embedded in the Hubble flow. Using this preferred direction~$\tensor*{\delta}{^\rho_0}$, there are only two tensor structures that can be present in~$\TT{^\mu_{\nu\sigma}}$, and these are parameterised by time-dependent scalar functions. The involved-looking normalisation of these scalars in~\cref{Tsamparlis}, and the shift involving the Hubble number, are purely for convenience in making sure that the modified Friedmann equations resulting from~\cref{TorCLagrangianMA} precisely match those of TorC. There is a formal way for deriving this general format for cosmological torsion using Lie derivatives~\cite{TSAMPARLIS197927} which, however, adds nothing of substance to the discussion.}
\begin{equation}\label{Tsamparlis}
		\tensor{\TT}{^\mu_{\nu\sigma}} 
		= \tensor*{\delta}{^\rho_0} \left[ (\phi + 2H)\tensor*{\delta}{^\mu_{[\sigma}}\tensor{g}{_{\nu]\rho}}
		- \sqrt{\frac{\Mp}{-3 \mu}}\varpi \tensor{\epsilon}{^\mu_{\rho\nu\sigma}} \right],
\end{equation}
where~$H(t)\equiv(\mathrm{d}a(t)/\mathrm{d}t)/a(t)$ is the Hubble parameter,~$\tensor{g}{_{\mu\nu}}$ is the metric tensor and~$\tensor{\epsilon}{^\mu_{\rho\nu\sigma}}$ is the totally antisymmetric tensor. A notable feature of the TorC model is that regardless of the value of the torsion scalar field~$\varpi$ in the early Universe, the field equations of TorC drive the scalar field towards unity~$\varpi \rightarrow 1$ as time progresses.\footnote{The~$\varpi$ field approaches the convenient value of unity due to the way in which it is normalised.} This is an attractor state in the dynamical system of the modified Friedmann equations, similar to how expanding de Sitter space is an attractor in GR when a cosmological constant is added. When the initial value of~$\varpi$ is set to be unity, corresponding to the condensed phase, the modified Friedmann equations of TorC become identical to those of GR at all future times, effectively reproducing the \LCDM{} model. The flexibility of this framework, however, allows for deviations from the \LCDM{} solution, effectively providing an extension to \LCDM{}. This offers a well-defined theoretical context for investigating possible resolutions to existing tensions in the \LCDM{} model, such as the Hubble tension. This work focuses on the background evolution of TorC cosmology and does not yet incorporate cosmological perturbations. For the purposes of this initial study, the perturbation equations remain those of standard \LCDM{}. This may seem an arbitrary choice, but for an initial investigation it provides a fiducial benchmark for future phenomenological development on the TorC research programme, as we will now explain.

\paragraph*{Fiducial perturbation theory}
In precision cosmology, perturbations are as vital as background dynamics, so our use of GR perturbation theory requires justification. TorC is currently treated as an inherently \emph{background} theory: while the Lagrangian in~\cref{eq:TorCLagrangian} was first proposed in~\cite{Lin:2019ugq} as having desirable perturbative features, later studies~\cite{Barker:2020gcp,Barker:2020elg,VandepeerBarker:2022xnp,Rew:2023zxy} showed that its non-perturbative background dynamics are both richer and shared by a wider model class. The bi-scalar-tensor formulation~\cref{TorCLagrangianMA}, where the 24 torsion components~$\TT{^\mu_{\nu\sigma}}$ reduce to two scalars~$\varpi$ and~$\phi$, reproduces the same background dynamics as~\cref{eq:TorCLagrangian}. An intermediate reduction between~\cref{TorCLagrangianMA,eq:TorCLagrangian} was presented in~\cite{VandepeerBarker:2022xnp}, where only the irreducible vector~$\TT{^\mu_{\nu\mu}}$ and axial-vector~$\epsilon_{\mu\nu\rho\sigma}\TT{^{\rho\sigma\mu}}$ parts of the torsion were retained.\footnote{As shown in~\cref{Tsamparlis},~$\phi$ and~$\varpi$ are essentially the time-aligned parts of these vectors, respectively.} This so-called \emph{tensor bypass}, in which the tensor part of the torsion is suppressed in the Lagrangian (see also~\cite{Barker:2022jsh,Barker:2023fem}), likewise reproduces the same background dynamics. Thus, TorC should be regarded as a research \emph{programme} rather than a single model,\footnote{In this sense, TorC may be compared to the research programme of modified Newtonian dynamics (MOND)~\cite{Mond,MondAQUAL}. The latter is not a specific theory, but a principle that may be realised through various specific theories~\cite{TeVeS,AeST}.} realisable with various field contents so long as the condensation principle is preserved. Different realisations may yield distinct perturbation theories on both the~$\TT \sim 0$ and~$\TT^2 \sim \Mp$ backgrounds. Hence, consistent model building requires tuning the field content--while keeping the condensation behaviour fixed--to make the~$\TT^2 \sim \Mp$ background a perturbatively unitary infrared foundation.\footnote{For example, we discussed above how any accidental symmetries of~\cref{eq:TorCLagrangian} may spoil the perturbation theory. The tensor bypass~\cite{VandepeerBarker:2022xnp} and bi-scalar-tensor~\cite{Barker:2020elg} realisations of TorC are expected to have \emph{fewer} such symmetries because, as previously mentioned, the gauge group of the quadratic limit of~\cref{eq:TorCLagrangian} near~$\TT\sim 0$ is mostly accounted for by the non-participation of the tensor parts of the torsion anyway.} To remain consistent with Newtonian and Minkowskian limits, perturbations near the condensate should not deviate strongly from GR. Accordingly, we study the background dynamics in isolation, adopting GR perturbation theory as a fiducial proxy.

\begin{table}[t]
\centering
\vspace{3mm}
\renewcommand{\arraystretch}{1.2} 
\setlength{\tabcolsep}{6pt}
\begin{tabular}{c|l}
\toprule
PGT & Poincar\'e Gauge Theory\\
TorC & Torsion Condensation\\
CMB & Cosmic Microwave Background\\
\hline
SH0ES & Local~$H_0$ measurement from SH0ES (2020)~\cite{Riess:2020fzl}\\
Planck & Planck 2018 data release~\cite{Planck:2018vyg}\\
\botrule
\end{tabular}
\caption{A list of acronyms and the datasets used in this analysis.}
\end{table}
\paragraph*{In this paper} In this work, the cosmological implications of the TorC model are investigated, with particular attention given to its potential to alleviate the Hubble tension. TorC introduces two parameters beyond those in \LCDM{}: the early-Universe value of the torsion scalar field~$\varpi_\mathrm{r}$ and its velocity, which may be `packaged' into the bare dark energy density parameter~$\Omega_\Lambda$. To constrain the TorC cosmological parameters, the nested sampling algorithm \texttt{PolyChord}~\cite{Handley:2015fda} is employed, interfaced via \texttt{Cobaya}~\cite{Torrado:2020dgo} with cosmological likelihoods and a modified version of \texttt{CAMB}~\cite{Lewis:1999bs}. A combined analysis is carried out between the Planck 2018 CMB data~\cite{Planck:2018vyg} and the 2020 SH0ES supernova measurement~\cite{Riess:2020fzl}, assessing the degree of tension between the two datasets using the~$R$-statistic~\cite{Handley:2019wlz, Hergt:2021qlh, Ormondroyd:2023cze}.

\paragraph*{Organisation of this paper} The paper is organized into the following sections: In~\cref{sec:cosmparam}, the two additional TorC parameters are introduced, along with their impact on the Universe's expansion history and the CMB power spectrum. In~\cref{sec:method}, the methodology for constraining the TorC parameters is outlined, including a description of prior choices and the likelihoods used in the analysis. Bayesian model comparison between \LCDM{} and TorC is also discussed, along with the tension statistics employed to assess the tension between datasets within the two models. In~\cref{sec:result}, the results of the parameter estimation are presented and compared to those obtained under the \LCDM{} model, in order to assess the viability of TorC as an alternative framework and its potential to resolve existing cosmological tensions. Finally, in~\cref{sec:conclusion}, the results are summarized, and the implications of the study are discussed, along with potential future directions of research.

\section{Cosmology}\label{sec:cosmparam}

\begin{table}[t]
\centering
\vspace{3mm}
\renewcommand{\arraystretch}{1.2} 
\setlength{\tabcolsep}{6pt}
\begin{tabular}{c|l}
\toprule
~$\Omega_\Lambda$ & Bare dark energy density parameter\\
~$\varpi_\mathrm{r}$ & Torsion scalar field value in the early Universe\\
\hline
$h$ & Dimensionless Hubble parameter\\
$\Omega_\mathrm{b}$ & Baryon density parameter\\
$\Omega_\mathrm{c}$ & Cold dark matter density parameter\\
$\tau_\mathrm{reio}$ & Reionisation optical depth\\
$A_\mathrm{s}$ & Amplitude of primordial scalar power spectrum\\
$n_\mathrm{s}$ & Primordial scalar spectral index\\
\botrule
\end{tabular}
\caption{Cosmological parameters of the TorC model. The top two rows list the two additional parameters introduced by TorC, while the bottom rows list the standard cosmological parameters in \LCDM{}.}
\label{tab:TorCparams}
\end{table}
The cosmology of TorC is explored under the assumption of a homogeneous and isotropic Universe. As an initial investigation, the scope of this work is restricted to the background evolution of TorC cosmology and does not consider cosmological perturbations. The perturbation equations therefore remain those of standard \LCDM{} throughout. 

This section focuses on the two additional parameters introduced by the model:~$\Omega_\Lambda$, the bare dark energy density parameter, and~$\varpi_\mathrm{r}$, specifying the value of the torsion scalar field in the early Universe, the list of cosmological parameters for TorC is presented in~\cref{tab:TorCparams}. In this section, their origin will be outlined. $\varpi_\mathrm{r}$ is an initial condition required for solving the field equations of TorC. $\Omega_\Lambda$ arises from the field equations as well, but with a crucial distinction from the standard model: unlike in \LCDM{}, where $\Omega_\Lambda$ is fixed by the constraint that all density parameters sum to unity, the dynamic torsion terms in TorC's Friedmann equations break this relationship. Consequently, $\Omega_\Lambda$ becomes a free parameter that must be constrained by observation.  TorC can be treated as an extension of \LCDM{} by absorbing the effects of~$\Omega_\Lambda$ and~$\varpi_\mathrm{r}$ into an effective dark energy density and pressure. This reformulated dark energy is incorporated into a modified version of \texttt{CAMB} to investigate their impact on the CMB power spectrum.
 
\subsection{TorC cosmological parameters:~$\Omega_\Lambda$ and~$\varpi_\mathrm{r}$}
\label{sec:TorCparams}
Varying the scalar-tensor analogue of the TorC Lagrangian in~\cref{TorCLagrangianMA} with respect to the metric~$\tensor{g}{_{\mu\nu}}$ and scalar fields~$\varpi$ and~$\phi$ gives the modified Friedmann equations for TorC~\cite{Barker:2020elg}. The field equation of~$\phi$ reveals that it is not an independent variable, but can be expressed in terms of the other fields 
\begin{equation}
	\phi = \frac{2\bigl(\dot{a} \bigl(\varpi^2 -1\bigr) + a \varpi \dot{\varpi} \bigr) }{a \bigl(\varpi^2 -1\bigr)} ,
\end{equation}
where the derivative with respect to time is denoted by a dot, e.g.,~$\dot{a} \equiv \mathrm{d}a/\mathrm{d}t$. This property allows for the immediate algebraic elimination of~$\phi$ from the other field equations. The variation of the matter Lagrangian with respect to the metric functions yields the energy-momentum tensor of a perfect fluid, characterised by the energy density~$\rho$ and pressure~$P$, which includes contributions from the standard radiation and matter. We additionally absorb the effect of~$\Lambda$ into this stress-energy tensor, to account for dark energy. These quantities can be written in the form of density parameters as
\begin{subequations}
\begin{align}
		\rho & = 3 \Mp H_0^2 (\Omega_\mathrm{r} + \Omega_\mathrm{m} + \Omega_\Lambda), \\
		P & = 3 \Mp H_0^2 \left(\frac{\Omega_\mathrm{r}}{3} - \Omega_\Lambda\right),
\end{align}
\end{subequations}
where~$H_0$ is the Hubble parameter today,~$\Omega_\mathrm{r}$ and~$\Omega_\mathrm{m}$ are the density parameters of radiation and matter, and~$\Omega_\Lambda$ represents the `intrinsic' dark energy density parameter, or the `bare' dark energy density parameter.

The field equations with respect to the two independent components of metric give analogues of the two Friedmann equations of GR, relating the content of the Universe to its expansion,
 \begin{samepage}
\begin{subequations}
 \begin{align}
		 H^2 = & H_0^2 \frac{\Omega_\mathrm{r} a^{-4}+\Omega_\mathrm{m} a^{-3}+\Omega_\Lambda}{\varpi^2} \nonumber \\
 & - \frac{\dot{\varpi} \left(6 H \varpi + \frac{\left(1+3\varpi^2\right)\dot{\varpi}}{\varpi^2-1}\right)}{3 \varpi^2}, \label{eq:TorCF1}\\
		 H^2 + \dot{H}^2 = & - H_0^2\frac{\Omega_\mathrm{r} a^{-4} + \frac{1}{2}\Omega_\mathrm{m} a^{-3} - \Omega_\Lambda}{\varpi^2}\nonumber\\
 &- \frac{3 H \varpi \dot{\varpi} - \frac{(5+3 \varpi^2) \dot{\varpi} ^2}{\varpi -1} + 3 \varpi \ddot{\varpi}}{3 \varpi^2}, \label{eq:TorCF2}
	\end{align}
\end{subequations}
 \end{samepage}
The third equation is the equation of motion of the remaining torsion scalar field, describing the evolution of the torsion scalar field
\begin{align}
\ddot{\varpi} = & \frac{1}{\left(\varpi^2-1\right)\left(3 \varpi^2 +1\right)} \nonumber\\
&\times \Bigl( -3 \varpi \left(\varpi^2 -1\right)^2 \left(2 H^2 + \dot{H}\right) \nonumber\\
&+ 3 H \left(1 + 2 \varpi^2 - 3 \varpi^4\right) \dot{\varpi}\nonumber \\
& + 4 \varpi \dot{\varpi}^2 \Bigr).\label{eq:varpiEOM}
\end{align}
The analogue of the first Friedmann equation in~\cref{eq:TorCF1} is compared with the first Friedmann equation of the \LCDM{} model
\begin{equation}\label{eq:LCDMF1}
		H^2 = H_0^2 \left(\Omega_\mathrm{r} a^{-4} + \Omega_\mathrm{m} a^{-3} + \Omega_{\Lambda}\right).
\end{equation}
It is evident that the equations~\cref{eq:TorCF1} and~\cref{eq:LCDMF1} differ in two key ways: first, the density parameters are effectively divided by~$\varpi^2$ in~\cref{eq:TorCF1}; second, an additional term appears in~\cref{eq:TorCF1}, involving the scale factor, torsion scalar field, and their derivatives. In standard \LCDM{}, the dark energy parameter~$\Omega_{\Lambda}$ is a constant that can be determined from the density parameters of the other components at~$a=1$ and~$H = H_0$, giving rise to the intuition that the non-gravitational energy budget of the Universe sums to unity when normalized. However, this relationship does not hold for TorC due to the presence of the extra term, similar to how a substantial curvature density would modify the Friedmann equations in GR. As a result,~$\Omega_\Lambda$ cannot be inferred from the other density parameters: it becomes an independent parameter of the model.

 To determine the expansion history of the Universe and the evolution of the torsion scalar field, the field equations are solved numerically in conformal time~$\tau$ defined as
\begin{equation}
	\mathrm{d}t = a \mathrm{d}\tau,
\end{equation}
using power series initial conditions for~$\varpi$ and~$a$ in the early Universe. As the bare dark energy density parameter~$\Omega_\Lambda$ plays a crucial role in shaping the late-time expansion of the Universe, it leads to a modified value of the current Hubble parameter~$H_0$. However, since~$H_0$ is a cosmological parameter fitted in the model, the scale factor~$a$ is normalized to ensure that the Hubble parameter reaches~$H_0$ at~$a=1$. This is implemented via a rescaling
\begin{equation}
	a \rightarrow \alpha a,
	\label{eq:rescale}
\end{equation}
where~$\alpha$ is a constant to be determined numerically.

The power series in the early Universe in~\cref{sec:TorCpower} reveals that the initial conditions for solving the system are functions of the early-Universe value of the torsion scalar field~$\varpi$, denoted~$\varpi_\mathrm{r}$. In addition to~$\Omega_\Lambda$,~$\varpi_\mathrm{r}$ thus emerges as an independent parameter of the model, playing a crucial role in early-Universe cosmology. The field equations of TorC thus introduce two new parameters,~$\Omega_\Lambda$ and~$\varpi_\mathrm{r}$, which influence the expansion history of the TorC Universe.

\subsection{Effective dark energy density and pressure}\label{sec:effectiveDE}
As discussed in~\cref{sec:intro}, TorC reproduces the \LCDM{} model when~$\varpi$ is set to unity. In this limit, the derivatives of~$\varpi$ vanish, and the TorC field equations reduce to the Friedmann equations of \LCDM{}. Therefore, if one wishes to reproduce \LCDM{} throughout the whole history of the Universe, one only needs to set the initial value of~$\varpi$ to unity,~$\varpi_\mathrm{r} = 1$, and the bare dark energy density parameter~$\Omega_\Lambda$ to its correspondent value in \LCDM{} model,~$\Omega_\Lambda = \Omega^\mathrm{\Lambda CDM}_\Lambda$. As \LCDM{} emerges as a particular dynamical evolution, based on a choice of initial conditions of the TorC model, TorC can be treated as an extension of \LCDM{}. The effect of the two additional parameters in TorC can be interpreted as deviations from the dark energy component of \LCDM{}. Rather than treating dark energy as a constant, it can be redefined as a time-dependent quantity, allowing the effects of the additional TorC terms to be absorbed into an evolving dark energy density and pressure. This reformulation enables the use of Boltzmann codes, such as \texttt{CAMB}, which are capable of incorporating modified dark energy models to investigate the impact of new parameters on the CMB power spectra. In this work, the treatment of dark energy is restricted to the background level, with effects of perturbation assumed to follow the standard \LCDM{}. The modified first Friedmann equation reflecting a time-dependent dark energy component takes the form
\begin{equation}
		H^2 = H_0^2 \bigl(\Omega_\mathrm{r} a^{-4} + \Omega_\mathrm{m} a^{-3}\bigr) +\frac{\rho_{\Lambda}^{\mathrm{eff}}(t)}{3 \Mp}.
	\label{eq:F1eff}
\end{equation}
By comparing~\cref{eq:F1eff} with~\cref{eq:TorCF1}, the additional terms in the TorC field equations can be reinterpreted as an effective dark energy density.
The effective dark energy density can be expressed as
 \begin{align}
		 \rho_{\Lambda}^{\mathrm{eff}} \equiv \frac{3\Mp}{\varpi^2} \Biggl[ &H_0^2 \Omega_\Lambda - H_0^2\left(\varpi^2 -1\right)\left(\Omega_\mathrm{r} a^{-4}+\Omega_\mathrm{m} a^{-3}\right) \nonumber\\
 & - 2 H \varpi \dot{\varpi} -\frac{\left(1+3 \varpi^2\right) \dot{\varpi}^2}{3\left(\varpi^2-1\right)}\Biggr].\label{eq:effrho}
\end{align}
Note that since the density parameters in~\cref{eq:TorCF1} are divided by~$\varpi^2$, they also appear in the effective dark energy density, as they cannot be isolated independently from~$\varpi$. The pressure of effective dark energy can then be obtained from the derivative of its density using the continuity equation\footnote{The assumption that effective fluid obeys a continuity equation follows precisely from the fact that the modified Friedmann equations are partitioned into a standard GR-like part and an effective modified gravity part. The continuity equation may then be derived purely due to the structure of the GR-like part: in turn this is a result of the Bianchi identity in GR, which forces the conservation of the stress-energy tensor to which gravity is coupled.}
\begin{equation}
	\dot{\rho} = -3 H \left(\rho + P\right).
\end{equation}
We thus obtain the following expression for the effective dark energy pressure
\begin{samepage}
\begin{align}
	P_{\Lambda}^\mathrm{eff} =& \frac{\Mp}{12 a^4 H \varpi^3(\varpi^2-1)}\Biggl[ 3 H_0^2 H \varpi \left(\varpi^2-1\right)\nonumber \\
							  & \times \Bigl(\Omega_\mathrm{r} \bigl(\varpi^2 -1\bigr) - 3 a^4 \Omega_\Lambda \bigl(1 + 3\varpi^2 \bigr) \Bigr) \nonumber\\
	& - 9 a^4 H^3 \varpi^3 \bigl(\varpi^2 -1\bigr)^2 \nonumber\\
	& + 6 \bigl(\varpi^2 -1\bigr) \dot{\varpi} \biggl(4 H_0^2 \bigl(\Omega_\mathrm{r} + \Omega_\mathrm{m} a + \Omega_\Lambda a^4 \bigr) \nonumber\\
	& -a^4 H^2 \varpi^2 \bigl(5 + 3 \varpi^2 \bigr)\biggr) \nonumber\\
	& - 3 a^4 H \varpi \dot{\varpi}^2 \bigl(-13 + 3\varpi^2 \bigl(6 + \varpi^2\bigr)\bigr)\nonumber\\	
	& - 8 a^4 \bigl(1 + 3\varpi^2\bigr) \dot{\varpi}^3 \Biggr].\label{eq:effP}
\end{align}
\end{samepage}
The effect of~$\Omega_\Lambda$ and~$\varpi_\mathrm{r}$ are thus absorbed into the effective dark energy density and pressure as a modification of time-dependent dark energy. By combining~\cref{eq:TorCF2,eq:varpiEOM}, one obtains a coupled system that governs the evolution of the torsion scalar field~$\varpi$ and the Hubble parameter. The equation of motion for~$\varpi$ determines how the scalar field evolves with time, while the modified Friedmann equation evolves the Hubble parameter, with each depending on the time-dependent values of the other. Together, these form an autonomous and self-consistent dynamical system: given suitable initial conditions, they can be solved simultaneously as a system of ordinary differential equations. This provides the evolution of~$\varpi$, its derivative, the Hubble parameter, and the scale factor. These solutions can then be substituted into the expressions for the time-dependent dark energy density and pressure in~\cref{eq:effrho,eq:effP}, yielding a complete description of cosmological evolution in the TorC framework. This allows for the exploration of their effects on the expansion history of the Universe and the CMB power spectra.

\subsection{Cosmological effects of~$\Omega_\Lambda$ and~$\varpi_\mathrm{r}$}
\label{sec:effects}
\begin{figure*}[!htbp]
 \centering
	\includegraphics[width=\textwidth]{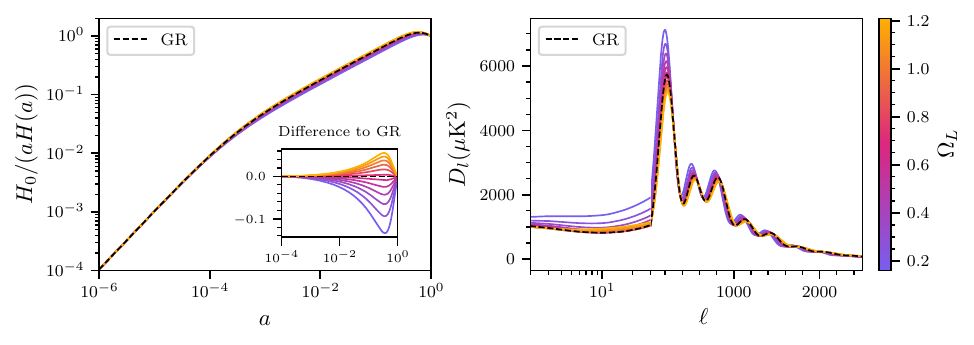}
	\caption{The effect of~$\Omega_\Lambda$ on the comoving Hubble horizon on the left and the CMB power spectrum on the right. The color bar indicates the value of~$\Omega_\Lambda$ in both panels. For both plots, the value for~$\varpi_\mathrm{r} \sim 1$ is set as the \LCDM{}-equivalent case. The cosmological parameters for computing CMB power spectra are set as the \LCDM{} values based on Planck results~\cite{Planck:2018vyg}. For comparison, the case for GR (based on~$\Omega_\mathrm{r} = 9.22 \times 10^{-5}$,~$\Omega_\mathrm{m} = 0.314$ and~$H_0 = 67.4$ km/s/Mpc) is shown in the plots. A smaller value of~$\Omega_\Lambda$ results in a smaller comoving Hubble horizon at late matter-dominated epochs, increases the amplitude of the CMB power spectrum, and shifts the peaks to lower multipoles. }
 \label{fig:OmLPlots}
\end{figure*}
\begin{figure*}[!htbp]
 \centering
	\includegraphics[width=\textwidth]{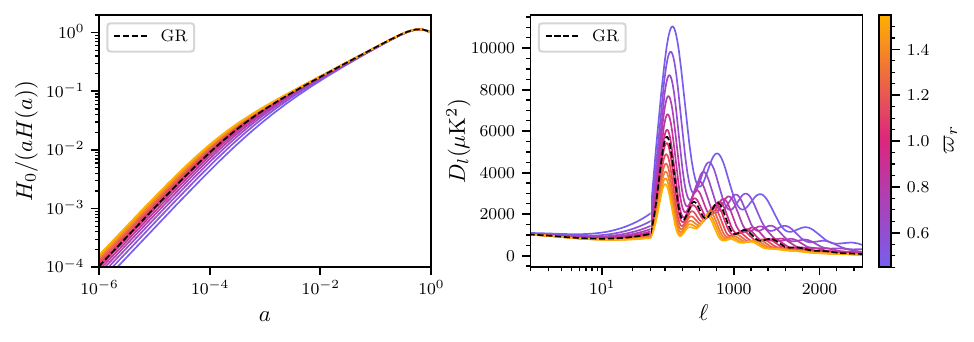}
	\caption{The effect of~$\varpi_\mathrm{r}$ on the Comoving Hubble Horizon on the left and the CMB power spectrum on the right. The color bar indicates the value of~$\varpi_\mathrm{r}$ in both panels. For both plots, the value for~$\Omega_\Lambda \sim 0.685$ is set as the \LCDM{}-equivalent case~\cite{Planck:2018vyg}. The other cosmological parameters are set as the values of the \LCDM{} model. For comparison, the case for GR (based on~$\Omega_\mathrm{r} = 9.22 \times 10^{-5}$,~$\Omega_\mathrm{m} = 0.314$ and~$H_0 = 67.4$ km/s/Mpc) is shown in the plots. A smaller value of~$\varpi_\mathrm{r}$ results in a smaller comoving Hubble horizon at early radiation-dominated epoch, increases the amplitude of the CMB power spectrum, and shifts the peaks to higher multipoles. }
 \label{fig:VarpirPlots}
\end{figure*}
To investigate the impact of the new TorC parameters on the CMB power spectrum, the model is treated as an extension of \LCDM{} and implemented with the \texttt{CAMB} Boltzmann code.  \texttt{CAMB} computes the CMB power spectra for a given cosmological model and supports a range of dark energy scenarios by allowing modifications to the dark energy equation of state $w_\Lambda^\mathrm{eff} \equiv P_\Lambda^\mathrm{eff}/\rho_\Lambda^\mathrm{eff}$.  However, attempting to incorporate the TorC extension through the equation of state $w_\Lambda^\mathrm{eff}(a)$ introduces poles when the dark energy density changes sign, as when~$\varpi$ exceeds unity in~\cref{eq:effrho}, the second and fourth terms become negative, potentially leading to a negative dark energy density during certain parts of the Universe’s evolution. This obstructs exploration of the full parameter space of~$\Omega_\Lambda$ and~$\varpi_\mathrm{r}$. However, the issue arises only when using $w_\Lambda^\mathrm{eff}$, and is resolved when pressure and density are implemented separately. To address this, \texttt{CAMB} was modified to evolve the system using the effective dark energy density and pressure directly.\footnote{Modified \texttt{CAMB} to read external dark energy density and pressure:~\cite{ModifiedCAMB}.}

\paragraph*{Effects of bare dark energy}~\cref{fig:OmLPlots} shows the cosmological effects of~$\Omega_\Lambda$. Note that this procesure --- adjusting bare dark energy density while holding the other parameters fixed --- is not possible in \LCDM{} as the density parameters have to sum up to unity. The left panel demonstrates that~$\Omega_\Lambda$ primarily influences the late-time expansion, with its effect most significant at late matter-dominated epochs, where a lower value of~$\Omega_\Lambda$ leads to a smaller comoving Hubble horizon.  Its contribution to the Hubble horizon is negligible at early times, resulting in an evolution of the early Universe that closely resembles \LCDM{}. The right panel of~\cref{fig:OmLPlots} shows that a decrease in~$\Omega_\Lambda$ shifts the CMB acoustic peaks to lower multipoles ($\ell$) and enhances their amplitudes. A lower~$\Omega_\Lambda$ causes photon-electron decoupling to occur at higher redshifts.  This results in a reduction in both the sound horizon,~$r^*$, and the angular diameter distance to the last scattering surface,~$D_A^*$.  However, a more significant reduction in~$D_A^*$ leads to an increase in the angular scale of the sound horizon, defined as~$\theta_s = r^*/D_A^*$, thereby shifting the acoustic peaks toward lower~$\ell$ values.  Additionally, the earlier decoupling prolongs the period of gravitational potential decay during the radiation-dominated era. This enhanced decay, via the early integrated Sachs-Wolfe (ISW) effect, injects additional energy into the photon-baryon fluid, amplifying the acoustic peak amplitudes.

\paragraph*{Effects of early torsion} The effects of~$\varpi_\mathrm{r}$ on the Hubble horizon and CMB power spectrum are shown in~\cref{fig:VarpirPlots}. The left panel shows that a decreasing~$\varpi_\mathrm{r}$ reduces the comoving Hubble horizon in the early Universe, while converging with \LCDM{} at late times. Correspondingly, the right panel shows that a smaller~$\varpi_\mathrm{r}$ shifts the acoustic peaks to higher multipoles~$\ell$ and increases their amplitudes. This shift results from the reduced sound horizon,~$r^*$ at decoupling, which represents the maximum distance that sound waves could travel through the primordial photon-baryon plasma before recombination. A higher expansion rate, or a smaller Hubble horizon at early times, means the universe expanded and cooled to the point of recombination more quickly. This results in a smaller sound horizon, leading to a higher angular scale of the sound horizon,~$\theta_s = r^*/D_A^*$, and thus shifting the acoustic peaks to higher multipoles~$\ell$. This mechanism is phenomenologically similar to that of early dark energy models (EDE)~\cite{Poulin:2018cxd, Kamionkowski:2022pkx}. However, unlike many EDE models, which introduce a new scalar field and potential for phenomenological gain, the torsional field in TorC arises uniquely from the unitary and power-counting renormalisable gauging of the Poincar\'e group, providing a more fundamental motivation. Furthermore, the smaller early Hubble horizon intensifies gravitational potential decay during the radiation-dominated era, thereby enhancing the early ISW effect, which acts as an additional driving force for photon-baryon oscillations and boosts the peak amplitudes. This modification to early Universe expansion also impacts Big Bang nucleosynthesis (BBN) and the resulting primordial element abundances, a detailed discussion of these effects is presented in~\cref{sec:Neff}.

\paragraph*{Mechanistic interpretability}
\begin{figure}[!htbp]
\includegraphics[width=\linewidth]{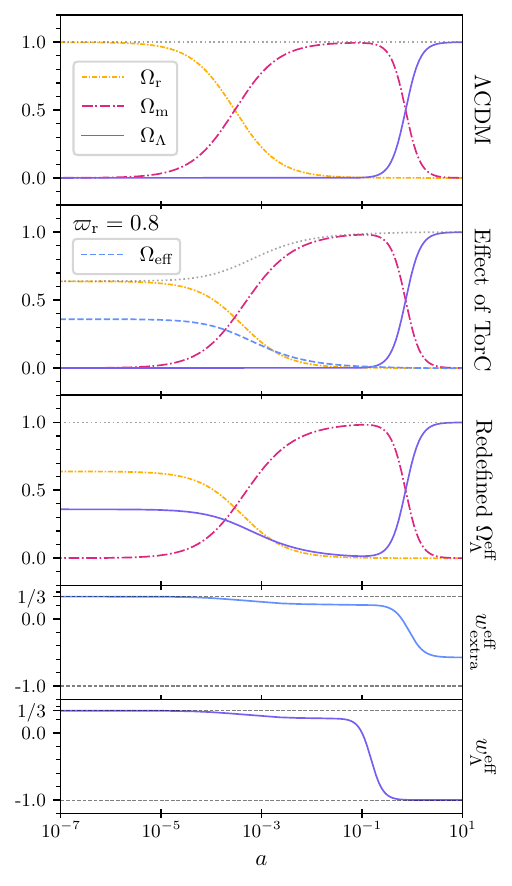}
\caption{Evolution of density parameters and effective equations of state in \LCDM{} and TorC cosmologies. The top panel shows the standard \LCDM{} components: radiation (yellow), matter (red), and dark energy (purple). The second panel shows the corresponding quantities in TorC cosmology. The additional TorC effect appears as an extra component (blue), representing the torsion contribution to the total energy budget of the Universe. A grey dotted line marks the combined contribution of the standard components (radiation, matter, and dark energy). The third panel displays the evolution of the effective dark energy density parameter in TorC, combining the bare dark energy and torsion contributions. The fourth and fifth panels show the equations of state for the additional extra torsional component and effective dark energy component, respectively. Both exhibit radiation-like behaviour ($w=1/3$) at early times, while the effective dark energy transitions to a cosmological-constant-like behaviour ($w=-1$) at late times. Since we focus on the effect of~$\varpi_\mathrm{r}$, no rescaling of the scale factor~$a$ is applied between models. TorC parameters are set to~$\Omega_\Lambda = 0.685$ and~$\varpi_\mathrm{r} = 0.8$, and \LCDM{} parameters follow the Planck 2018 results~\cite{Planck:2018vyg}.}
\label{fig:densityevolution}
\end{figure}
\cref{fig:densityevolution} illustrates the evolution of the density parameters in \LCDM{} and TorC cosmologies. The top panel shows the familiar evolution in \LCDM{}, highlighting the three density-dominated epochs. The second panel introduces the TorC modification, expressed as an additional component that can be interpreted as the torsion contribution to the total energy budget of the Universe. This extra component contributes primarily at early times, consistent with the early Universe modification shown in~\cref{fig:VarpirPlots}. The third panel shows the evolution of the effective dark energy density parameter in TorC, which incorporates both the torsional contribution and bare dark energy. This demonstrates how the effect of TorC parameters being absorbed into a redefined effective dark energy, corresponding to the density parameter in~\cref{eq:effrho}. The bottom two panels display the equations of state for the additional extra component and the effective dark energy component, respectively. The fourth panel shows that the TorC extra component behaves like radiation at early times, with an equation of state $w^\mathrm{eff}_\mathrm{extra}=1/3$, before freezing out to a different constant value during the matter and dark energy dominated epochs (see~\cite{Barker:2020gcp} for further discussion). Finally, the bottom panel shows that the effective dark energy component, which combines effects from $\varpi_\mathrm{r}$ and the bare dark energy, evolves from a radiation-like equation of state ($w^\mathrm{eff}_\Lambda=1/3$) at early times to a cosmological-constant-like behaviour ($w^\mathrm{eff}_\Lambda=-1$) at late times. This dynamical evolution drives the modifications observed in the Hubble horizon and CMB power spectra. A related discussion of the effect of~$\Omega_\Lambda$ is given in~\cref{sec:matter-evolution}, where it is shown to shift the epochs of density equivalence. Note that the effective dark energy equation of state exhibits poles, motivating our use of effective density and pressure rather than $w^\mathrm{eff}_\Lambda$ itself in this work.

\paragraph*{Connection to the Hubble tension}
An approach for addressing the Hubble tension is to modify the early-time expansion rate so that the CMB-inferred value of~$H_0$ shifts upward toward local measurements such as SH0ES. Any such modification, however, must preserve the precisely measured angular scale of the sound horizon,~$\theta_s = r^*/D_A^*$, which is tightly constrained by the acoustic peaks in the CMB power spectrum. In the TorC model, enhancing the early expansion rate (i.e.~$\varpi_r < 1$) reduces the time available for sound waves to propagate before recombination, resulting in a smaller physical sound horizon,~$r^*$. To maintain the fixed observational ratio~$\theta_s = r^*/D_A^*$, the comoving angular diameter distance to the last scattering surface,~$D_A^*$, must also decrease proportionally. This distance depends on the expansion history since recombination and is inversely proportional to the Hubble constant. For a general curvature~$\Omega_\mathrm{k}$,\footnote{Note that when $\Omega_\mathrm{k} > 0$, the argument of the sine becomes imaginary, and the identity $\sin(ix) = i\,\sinh(x)$ converts the sine into a hyperbolic sine. In the limit $\Omega_\mathrm{k} \to 0$, the expression smoothly reduces to the flat-universe result, so this form consistently covers all curvature cases.} it is given by~\cite{Hogg:1999ad, Barker:2020gcp}
\begin{equation}
D_A(z*) \equiv \frac{\sin \Bigl(\sqrt{-\Omega_\mathrm{k}} \int^{z*}_0 \frac{H_0 \mathrm{d} z}{H(z)} \Bigr)}{H_0 \sqrt{-\Omega_\mathrm{k}}},
\end{equation}
where $z*$ is the redshift at recombination. A lower~$D_A^*$ therefore demands a larger inferred~$H_0$. In this way, the TorC model naturally shifts the CMB-inferred Hubble constant to a higher value, bringing it into closer agreement with local measurements and thereby alleviating the tension. This general mechanism is shared by many other proposed solutions that modify early-universe physics, such as early dark energy~\cite{Mortsell:2018mfj, Poulin:2018cxd, Kamionkowski:2022pkx} or varying dark energy models~\cite{Elgaroy:2007bv}. The TorC model has the advantage that its modifications arise from a well-motivated theoretical framework based on gauging the Poincar\'e group, offering a more fundamental origin for the new physics.

\section{Methods}\label{sec:method}

The previous sections introduced the two new parameters in TorC,~$\Omega_\Lambda$ and~$\varpi_\mathrm{r}$, and their effects on the Hubble horizon and CMB power spectra of the TorC Universe. To find the posterior distribution of the cosmological parameters in TorC, the \texttt{Polychord}~\cite{Handley:2015vkr} nested sampling algorithm is employed to explore the parameter space of TorC cosmological parameters. The sampling is conducted using \texttt{Cobaya}~\cite{Torrado:2020dgo} which interfaces the \texttt{CAMB} Boltzmann solver with \texttt{Polychord} and relevant cosmological likelihoods. Note that this analysis modifies the background expansion in \texttt{CAMB} using the TorC effective dark energy density and pressure, while the perturbation equations remain those of standard \LCDM{}.\footnote{As a initial study, only the background evolution of TorC is being investigated, cosmological perturbation theory for TorC will be developed in future work.}

The two models, TorC and \LCDM{}, are compared using the Bayesian evidence,~$\mathcal{Z}$. The tension between the Planck and SH0ES datasets is then assessed using the~$R$-statistic, which quantifies the degree of tension and its potential alleviation when comparing the \LCDM{} and TorC models.

\subsection{Numerical implementation}

As discussed in~\cref{sec:effects}, to explore the full parameter space of $\Omega_\Lambda$ and $\varpi_\mathrm{r}$, the \texttt{CAMB} Boltzmann code is modified to read in a tabulated input of dark energy density~$\rho_\Lambda(a)$, and pressure~$P_\Lambda(a)$ as functions of scale factor~$a$. The modified version is available at~\cite{ModifiedCAMB}. This customized \texttt{CAMB} is then linked to a modified \texttt{Cobaya} which reads the same external~$\rho_\Lambda(a)$ and~$P_\Lambda(a)$, available at~\cite{ModifiedCobaya}. 
The chains and supplementary files used to produce the plots and results presented in this paper are available at~\cite{ZenodoTorC}. 

\subsection{Prior and likelihoods}
\label{sec:prior}
\begin{table}[t]
\centering
\vspace{3mm}
\renewcommand{\arraystretch}{1.2}
%\begin{tabularx}{\linewidth}{c|c}
\begin{tabularx}{\linewidth}{>{\centering\arraybackslash}X|>{\centering\arraybackslash}X}
\toprule
Parameter & Prior Range \\
\hline
~$\Omega_\Lambda$ &~$[0.1, 1.5]$ \\
~$\varpi_\mathrm{r}$ &~$[0.1, 1.5]$ \\
\hline
~$h$ &~$[0.2, 1.0]$ \\
~$\Omega_\mathrm{b} h^2$ &~$[0.005, 0.1]$ \\
~$\Omega_\mathrm{c} h^2$ &~$[0.005, 0.99]$ \\
~$\tau_{\mathrm{reio}}$ &~$[0.01, 0.4]$ \\
~$n_\mathrm{s}$ &~$[0.885, 1.04]$ \\
~$\log A_\mathrm{s}$ &~$[2.5, 3.7]$ \\
\botrule
\end{tabularx}
\caption{Prior range for uniform priors on cosmological parameters in the TorC model and \LCDM{} model. The prior range for the \LCDM{} parameters comes from the default option in~\texttt{cobaya}.} 
 \label{tab:prior}
\end{table}
To explore the parameter space of the TorC model, uniform priors are adopted for the set of parameters~$\{\Omega_\Lambda$,~$\varpi_\mathrm{r}$,~$h$,~$\Omega_\mathrm{b} h^2$,~$\Omega_\mathrm{c} h^2$,~$\tau_{\mathrm{reio}}$,~$n_\mathrm{s}$,~$A_\mathrm{s}\}$.~\footnote{The prior of~$\varpi_\mathrm{r}$ is restricted to avoid singularity at~$\varpi_\mathrm{r} = 1$, and a prior constraint is imposed on the total matter density, requiring~$\Omega_\mathrm{m} = \Omega_\mathrm{b} + \Omega_\mathrm{c} < 1$ to avoid segmentation faults.} The last six parameters correspond to the standard cosmological parameters in the \LCDM{} model.  The prior ranges for the cosmological parameters are summarized in~\cref{tab:prior}.

The likelihoods used in this study include the Planck 2018 high-$\ell$ and low-$\ell$ TT, TE, EE, and lensing likelihoods~\cite{Aghanim:2019ame}.  Additionally, the 2020 SH0ES measurement of the present-day Hubble parameter,~$H_0 = 73.04 \pm 1.04$ \hunit~\cite{Riess:2020fzl}, is also included via a likelihood with a Gaussian profile in the parameter~$H_0$, enabling comparison between the two datasets and combined analysis.  These likelihoods and priors are applied to both the TorC and \LCDM{} analysis.

\subsection{Bayesian model comparison}
Bayesian model comparison between TorC and \LCDM{} is conducted using the Bayesian evidence
 \begin{equation}
 		\mathcal{Z} = \int \mathcal{L}(\theta) \pi(\theta) \mathrm{d}\theta,
 \end{equation}
 where~$\mathcal{L}(\theta)$ is the likelihood function given the model parameters~$\theta$, and~$\pi(\theta)$ is the prior probabilities. The evidence naturally incorporates `\emph{Occam's razor}', penalizing overly complex models. To better understand this, the evidence can be decomposed into two components in the logarithmic form~\cite{Handley:2015vkr, Balan:2024cmq}
\begin{equation}
		\log \mathcal{Z} = \langle \log \mathcal{L}(\theta) \rangle_\mathcal{P}- \mathcal{D}_{\mathrm{KL}}.
\end{equation}
The first term is the posterior-weighted log likelihood,
\begin{equation}
		\langle \log \mathcal{L}(\theta) \rangle_\mathcal{P} = \int \log \mathcal{L}(\theta) P(\theta | D) \mathrm{d}\theta,
\end{equation}
where~$P(\theta | D)$ is the posterior distribution of the parameters given the data~$D$. This term measures how well the model fits the data, averaged over the posterior distribution of the parameters. A larger value of this indicates that the model provides a better fit to the data.
The second term is the Kullback-Leibler (KL) divergence~\cite{Handley:2019pqx, Kullback:1951zyt}
\begin{equation}
		\mathcal{D}_{\mathrm{KL}} = \int P(\theta | D) \log \frac{P(\theta | D)}{\pi(\theta)} \mathrm{d}\theta,
\end{equation}
which quantifies how much the posterior distribution has contracted relative to the prior. This term acts as a complexity penalty --- a model with a large parameter space that is significantly narrowed by the data will have a larger value of~$\mathcal{D}_{\mathrm{KL}}$.

For models~$M_1$ and~$M_2$, model comparison can be conducted using Bayes' theorem. Given uniform prior probabilities for the two models, the posterior probability~$M_1$ is given by
\begin{equation}
		P(M_1 | D) = \frac{\mathcal{Z}_1}{\mathcal{Z}_1 + \mathcal{Z}_2}.
\end{equation}
The ratio of posterior probability for the two models,~$P(M_1 | D) / P(M_2 | D) = \mathcal{Z}_1 / \mathcal{Z}_2$, gives Bayes factor, which offers a direct measurement of the relative support for the two models, quantifying how much more strongly the data favors~$M_1$ over~$M_2$. In the logarithm form, the Bayes factor is given by
\begin{equation}
		\begin{split}
				\Delta \log \mathcal{Z} & = \log \mathcal{Z}_1 - \log \mathcal{Z}_2.\\
										& = \Delta \langle \log \mathcal{L}(\theta) \rangle_\mathcal{P} - \Delta \mathcal{D}_{\mathrm{KL}},
		\end{split}
		\label{eq:BayesFactor}
\end{equation}
values of~$\Delta \log \mathcal{Z} > 0$ indicate that $M_1$ is supported over $M_2$ by a factor of $\exp(\Delta \log \mathcal{Z})$, while~$\Delta \log \mathcal{Z} < 0$ suggests the opposite.
As shown in~\cref{eq:BayesFactor}, this framework allows the source of model preference to be identified, comparing the goodness-of-fit and the complexity penalty of the two models. This analysis will be performed between TorC and \LCDM{}, exploring the added complexity of the two new parameters~$\Omega_\Lambda$ and~$\varpi_\mathrm{r}$ in TorC, and their impact on the fit to the datasets.

\subsection{Tension quantification}
The level of agreement between the Planck and SH0ES datasets is assessed using the~$R$-statistic with \texttt{anesthetic}~\cite{Handley:2019mfs}, which quantifies the relative confidence in one dataset given the other, compared to the confidence in the dataset given itself. For datasets~$A$ and~$B$, the~$R$-statistic is defined as
\begin{equation}
 R = \frac{\mathcal{Z}_{AB}}{\mathcal{Z}_A \mathcal{Z}_B} = \frac{P(A, B)}{P(A)P(B)} = \frac{P(A|B)}{P(A)} = \frac{P(B|A)}{P(B)},
\end{equation}
where~$\mathcal{Z}_{AB}$ is the joint evidence for datasets~$A$ and~$B$, and~$\mathcal{Z}_A$ and~$\mathcal{Z}_B$ are the evidences for each dataset individually. For~$R \gg 1$, the two datasets are consistent, indicating that dataset~$B$ has strengthened our confidence in dataset~$A$ by a factor of~$R$. Conversely, for~$R \ll 1$, the two datasets are in tension.

For flat, uninformative priors, the~$R$-statistic becomes sensitive to prior volume. In such cases, increasing the prior range can artificially inflate~$R$, falsely suggesting greater agreement. This issue can be resolved by using the Bayesian suspiciousness metric~\cite{Handley:2019wlz}, which accounts for prior volume effects. In the present analysis, the~$R$-statistic is sufficient to assess the level of agreement between the two datasets.

\section{Results}\label{sec:result}

\begin{figure*}[!htbp]
	\centering
	\includegraphics[width=\textwidth]{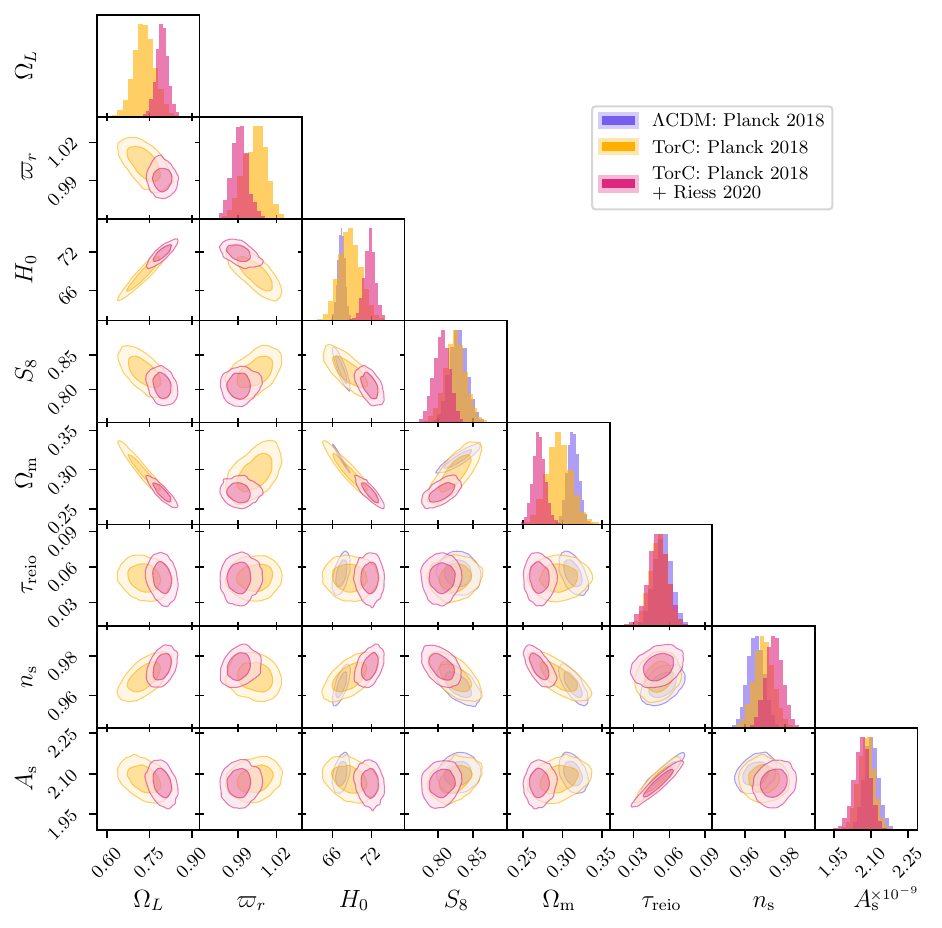}
	\caption{The figure shows the full posterior distributions of the cosmological parameters for \LCDM{} (Planck 2018 data (purple)) and TorC (Planck 2018 data (yellow)
and joint Planck+SH0ES data (red)). Nested sampling was performed with \texttt{PolyChord} using 1000 live points, and the corner plot was generated with \texttt{anesthetic}. } \label{fig:TorC_LCDM_posterior}
\end{figure*}

\cref{fig:TorC_LCDM_posterior} shows the full posterior distributions of the TorC and \LCDM{} cosmological parameters. the result is obtained from \texttt{PolyChord} nested sampling with 1000 live points. 

\begin{figure*}[!htbp]
\centering
\begin{subfigure}{0.49\textwidth}
\centering
\includegraphics[width=\textwidth]{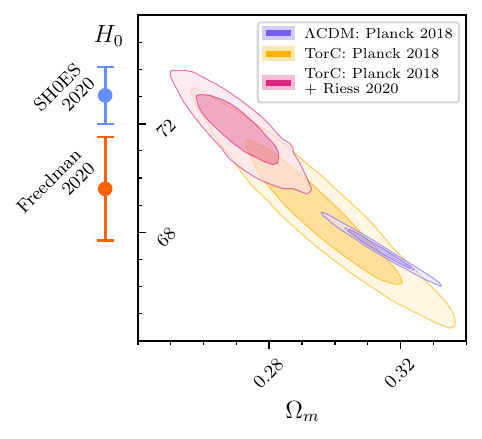}
\end{subfigure}
\hfill 
\begin{subfigure}{0.49\textwidth}
\centering
\includegraphics[width=\textwidth]{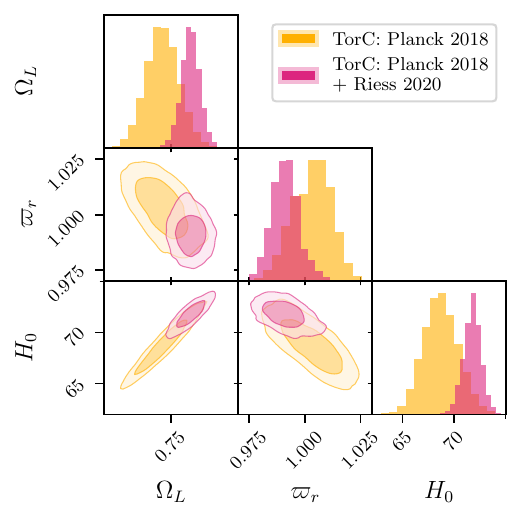}
\end{subfigure}
\caption{The left panel shows~$\Omega_\mathrm{m}$-$H_0$ posterior contours for \LCDM{} (using Planck 2018 data (purple), TRGB data (orange)~\cite{Freedman:2020dne}, and SH0ES 2020 data (blue)~\cite{Riess:2020fzl}) and for TorC (using Planck 2018 data (yellow), and joint Planck and SH0ES data (red)). The contours show the tension between the Planck and SH0ES posteriors in the \LCDM{} model, and the alleviation of this tension in the TorC model. The right panel shows the posterior distributions of the two TorC parameters~$\Omega_\Lambda$ and~$\varpi_\mathrm{r}$ in the~$H_0$-plane. The plot shows a positive correlation between~$H_0$ and~$\Omega_\Lambda$, and a negative correlation between~$H_0$ and~$\varpi_\mathrm{r}$. }
\label{fig:corners}
\end{figure*}

\begin{figure*}[!htbp]
\centering
\begin{subfigure}{0.49\textwidth}
\centering
\includegraphics[width=\textwidth]{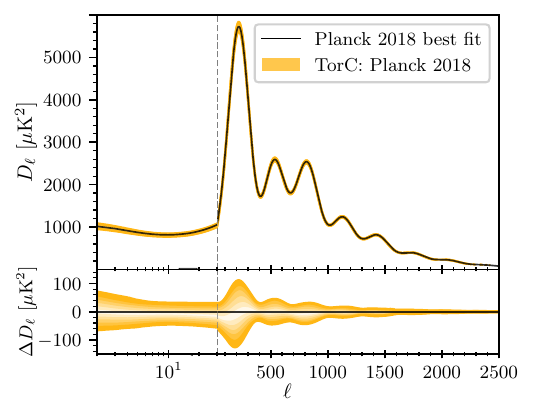}
\end{subfigure}
\hfill
\begin{subfigure}{0.49\textwidth}
\centering
\includegraphics[width=\textwidth]{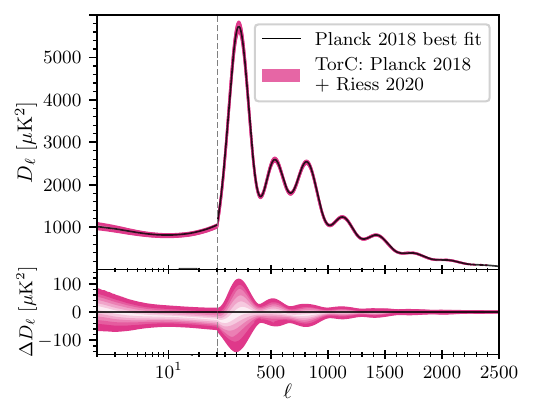}
\end{subfigure}
\caption{The posterior results of the CMB temperature power spectrum in TorC cosmology with both Planck data alone (left), and combined Planck and SH0ES data (right). The upper panels display the posterior distribution of the CMB power spectra, and the Planck 2018 best fit for \LCDM{} model (solid line), and the bottom panels show the residuals in comparison with the Planck 2018 best fit \LCDM{} model~\cite{Planck:2018vyg}. Both plots are generated with the respective posterior distributions using \texttt{fgivenx}~\cite{fgivenx}. }
\label{fig:cmb_residuals}
\end{figure*}

\subsection{$H_0$ Tension}
The TorC model is capable of alleviating the~$H_0$ tension by allowing for a higher inferred Hubble parameter compared to the \LCDM{} model. This is shown in the~$H_0$-$\Omega_\mathrm{m}$ plane on the left panel of~\cref{fig:corners}. The \LCDM{} posteriors with Planck, SH0ES, and tip of the red giant branch (TRGB)~\cite{Freedman:2020dne} likelihoods reveal a clear tension between the Planck and SH0ES datasets. Whereas, the TorC posteriors for both Planck likelihood alone and joint SH0ES and Planck likelihoods show a fit that is consistent with both datasets.

It is important to clarify the nature of this tension alleviation. A successful resolution is achieved when a physically motivated framework demonstrates the statistical consistency of both early- and late-universe probes. The introduction of the torsional degree of freedom in TorC, which behaves as an early dark radiation component as shown in~\cref{fig:densityevolution}, naturally broadens the posterior distribution for~$H_0$ derived from CMB data. While the peak of the posterior from a Planck-only analysis within TorC may remain lower than the SH0ES value, the tension is alleviated because the SH0ES measurement is no longer a significant statistical outlier --- it now falls within the broadened confidence region of the CMB constraint. This reduces what would be a~$\sim5\sigma$ discrepancy in \LCDM{} to a statistically acceptable level in TorC. The degree of this alleviation is quantified in~\cref{sec:tension} using the~$R$-statistic. This approach provides a genuine resolution by demonstrating the consistency of both datasets within a physically motivated framework.

The right panel of~\cref{fig:corners} shows the posterior distributions of the two TorC parameters,~$\Omega_\Lambda$ and~$\varpi_\mathrm{r}$, along with $H_0$. The~$H_0$-$\varpi_\mathrm{r}$ plane shows a negative correlation: as~$\varpi_\mathrm{r}$ decreases, the value of~$H_0$ increases. This behavior suggests that while the TorC model is consistent with \LCDM{}, it allows for a higher value of~$H_0$ through the presence of torsion. Furthermore, the~$H_0$-$\Omega_\Lambda$ plane shows a positive correlation, indicating that a higher~$H_0$ value in TorC is achieved not only through a reduction in~$\varpi_\mathrm{r}$, but also by an increase in the amount of bare dark energy~$\Omega_\Lambda$.

The posterior distributions of the CMB temperature power spectra in TorC cosmology, obtained using Planck data alone and using the combined Planck+SH0ES dataset, are shown in~\cref{fig:cmb_residuals}.The corresponding late-time evolution of the Hubble parameter is presented in~\cref{sec:hubble-evolution}, illustrating how the TorC posterior approaches $H_0$. The left panel of~\cref{fig:cmb_residuals} displays the Planck-only result, while the right panel presents the joint fit. In both panels, we also show the residuals relative to the Planck 2018 best-fit \LCDM{} spectrum. The residuals reveal a small shift in the acoustic peak structure, but the Planck best-fit \LCDM{} model remains well within the TorC posterior contours. This demonstrates that the TorC model provides a fit to the CMB data that is fully consistent with the standard model. Overall, the plots demonstrate that while TorC can accommodate a higher~$H_0$ value, its CMB power spectrum remains very close to that of \LCDM{}.

\subsection{On~$S_8$}
As shown in~\cref{fig:TorC_LCDM_posterior}, when including the Planck likelihood alone, the TorC posterior allows for lower values of~$S_8$ while being consistent with the \LCDM{} result. Including SH0ES likelihood shifts the TorC posterior nominally closer to the KiDS-1000 result,~$S_8 = 0.745 \pm 0.039$~\cite{KiDS:2020suj}. However, these results must be interpreted with caution, as the current analysis does not include TorC cosmological perturbations. It is also worth noting that the more recent KiDS-Legacy analysis shows that the~$S_8$ value is consistent with the result from Planck~\cite{Wright:2025xka}. In future work, the TorC model will be extended to include perturbation analysis, which may provide a more accurate prediction of~$S_8$.

\subsection{Model comparison}
\label{sec:model-comparison}
\begin{figure*}[!htbp]
    \centering
	\includegraphics[width=\textwidth]{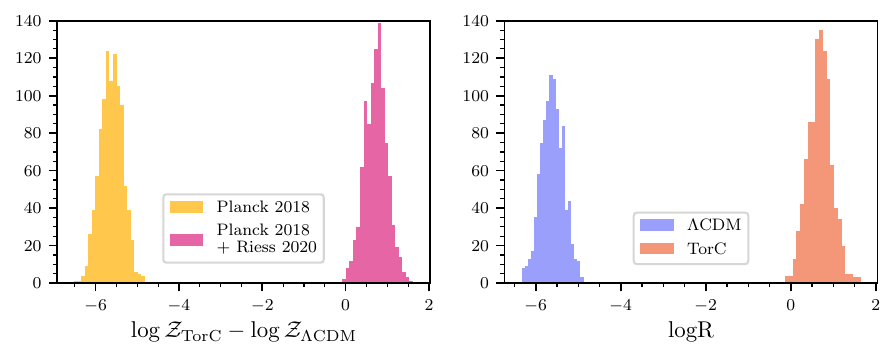}
	\caption{The left panel shows the logarithmic Bayes factor between \LCDM{} and TorC models using Planck data and the joint SH0ES and Planck data, defined as~$\Delta \log \mathcal{Z} = \log \mathcal{Z}_{\mathrm{TorC}} - \log \mathcal{Z}_{\LCDM{}}$.
	The right panel shows the logarithmic~$R$-statistic between Planck and SH0ES datasets for \LCDM{} and TorC models.
	In both cases the statistics are simulated with 1000 data points.
}
\label{fig:LogZLogR}
\end{figure*}
Bayesian model comparison between the TorC and \LCDM{} models was carried out using both the Planck data alone and the joint Planck and SH0ES datasets. The Bayes factors between the TorC and \LCDM{} models were computed using \texttt{PolyChord}, the results are shown in the left panel of~\cref{fig:LogZLogR}. The Bayes factor is defined as~$\Delta \log \mathcal{Z} = \log \mathcal{Z}_{\mathrm{TorC}} - \log \mathcal{Z}_{\LCDM}$ in this case.

Writing the logarithmic Bayes factor in the form of the second line of~\cref{eq:BayesFactor}, namely~$\log \Delta \mathcal{Z} = \Delta \langle \log \mathcal{L}(\theta) \rangle_\mathcal{P} - \Delta \mathcal{D}_\mathrm{KL}$, the result with Planck 2018 alone is
\begin{equation} \label{eq:LogZPlanck}
	\begin{split}
			\Delta \log \mathcal{Z}^{\mathrm{Planck}} &= -5.627 \pm 0.269 \\
									&= 0.417 \pm 0.114 - 6.044 \pm 0.265,
	\end{split}
\end{equation}
and with the joint Planck and SH0ES datasets is
\begin{equation} \label{eq:LogZPlanckSH0ES}
		\begin{split}
				\Delta \log \mathcal{Z}^\mathrm{joint} &= 0.725 \pm 0.280 \\
								&= 6.792 \pm 0.113 - 6.071 \pm 0.263.
		\end{split}
\end{equation}
In both cases, the relative KL divergence is essentially the same, reflecting the complexity penalty from the two additional parameters in TorC.~\cref{sec:TorCprior} illustrates this penalty by showing the additional parameter space in TorC and the compression of the prior into the posterior, which is quantified by the KL divergence term.

For the Planck-only analysis in~\cref{eq:LogZPlanck}, there is a positive contribution from the likelihood term, however, this improvement in fit is outweighed by the complexity penalty, resulting in a strong evidence favouring \LCDM{} over TorC. In the joint Planck and SH0ES analysis in~\cref{eq:LogZPlanckSH0ES}, the likelihood term becomes larger, indicating that TorC provides an improved fit to the joint datasets. However, this improvement does not sufficiently offset the complexity penalty, leading to a Bayesian evidence for TorC that is only comparable to \LCDM{}. This suggests while TorC can accommodate both datasets, its additional complexity does not provide a decisive advantage over \LCDM{}.

\subsection{Tension analysis between datasets}
\label{sec:tension}
The tension between the Planck and SH0ES datasets is quantified using the~$R$-statistic, and a comparison of~$\log R$ values between the \LCDM{} and TorC models is presented in the right panel of~\cref{fig:LogZLogR}. The \LCDM{} model shows a significant tension between the two datasets ($\log R = -5.60 \pm 0.28$), while the TorC model suggests consistency between them ($\log R = 0.70 \pm 0.27$). This improvement in the~$R$-statistic for the TorC model demonstrates its ability to increase compatibility between the Planck and SH0ES datasets, quantifying the alleviation of the~$H_0$ tension.

\section{Conclusions}\label{sec:conclusion}

This study investigates the torsion condensation (TorC) model --- a gravitational gauge field theory with quadratic Lagrangian terms in torsion and curvature. The TorC cosmology can be studied as an extension of the \LCDM{} model, introducing two new parameters,~$\Omega_\Lambda$ and~$\varpi_\mathrm{r}$, corresponding to the bare dark energy density and value of the torsional scalar field in the early Universe, respectively. The effects of these new parameters on the Hubble horizon and CMB power spectra are explored. In particular, the torsional degree of freedom can enhance the early-time expansion rate, leading to a higher inferred value of the Hubble constant~$H_0$ from CMB data. The constraints on the TorC cosmological parameters are then obtained using the \texttt{PolyChord} nested sampling algorithm, interfaced with the \texttt{CAMB} Boltzmann code and Planck and SH0ES likelihoods through \texttt{Cobaya}.~$H_0$ shows a negative correlation with~$\varpi_\mathrm{r}$ and a positive correlation with~$\Omega_\Lambda$, indicating that an introduction of torsion in the gravitational sector can increase the value of~$H_0$. Tension analysis using the~$R$-statistic quantifies the alleviation of the tension between the Planck and SH0ES datasets in the TorC model, further demonstrating the ability of TorC to increase compatibility between the two datasets. However, model comparison between TorC and \LCDM{} using Bayes factors shows that TorC is not decisively favoured against \LCDM{}: the improvement in fit does not significantly offset the complexity penalty from the two additional TorC parameters. Nevertheless, this study demonstrates the potential of the TorC model to address the~$H_0$ tension, and motivates further investigation into the cosmological implications of extended gravitational theories.

Future work will extend the analysis of TorC parameters by incorporating other cosmological likelihoods, such as the recent DESI BAO dataset~\cite{DESI:2025zgx}. The present study assumes a vanishing torsional dark energy component,~$\lambda = 0$, future investigations will consider scenarios in which both bare and torsional contributions to dark energy are present. The inclusion of torsional dark energy may offer a natural explanation for the remarkably small observed value of the cosmological constant; the concept of torsion acting as a dynamical source for dark energy is also an active field of research in other extended gravity frameworks~\cite{Benisty:2021sul, Benisty:2022lhr}. This study is also conducted in the context of a flat Universe. Future work will also explore scenarios with non-zero spatial curvature. While the TorC field equations for background evolution are insensitive to the spatial curvature constant~$k$, spatial curvature can still manifest through its standard geometric effects on light propagation and large-scale structure. It will be investigated whether TorC under non-zero spatial curvature can alleviate the curvature tension observed in the \LCDM{} model. Finally, the development of a perturbation theory for TorC will enable a deeper exploration of the model impact on cosmic structure formation and the CMB.

\begin{acknowledgments}
The authors would like to thank Mike Hobson, Anthony Lasenby, and Dily Ong for useful discussions. SL would like to thank the IBM colour blind friendly palette for the color scheme used in the figures, and Joe Pattison and Oscar O'Hara for their advice on making the plots more visually appealing. WB is grateful for the support of Girton College, Cambridge, Marie Skłodowska-Curie Actions and the Institute of Physics of the Czech Academy of Sciences.

This work used the DiRAC Data Intensive service~(CSD3 \href{www.csd3.cam.ac.uk}{www.csd3.cam.ac.uk}) at the University of Cambridge, managed by the University of Cambridge University Information Services on behalf of the STFC DiRAC HPC Facility~(\href{www.dirac.ac.uk}{www.dirac.ac.uk}). The DiRAC component of CSD3 at Cambridge was funded by BEIS, UKRI and STFC capital funding and STFC operations grants. DiRAC is part of the UKRI Digital Research Infrastructure. The authors acknowledge the use of \texttt{NumPy}~\cite{harris2020array}, \texttt{SciPy}~\cite{2020SciPy-NMeth}, and \texttt{Matplotlib}~\cite{Hunter:2007} for numerical computations and plotting, and \texttt{gemini-2.5-flash-preview-04-17} for the improvement of clarity of this paper.

\leavevmode

\paragraph*{Disclaimer} Co-funded by the European Union (Physics for Future --- Grant Agreement No. 101081515). Views and opinions expressed are however those of the author(s) only and do not necessarily reflect those of the European Union or European Research Executive Agency. Neither the European Union nor the granting authority can be held responsible for them.
\end{acknowledgments}

\bibliography{NotINSPIRE,paper_Qtorsion}

\appendix

\section{Condensation and perturbation}\label{sec:perturbations}

\paragraph*{Poincar\'e gauge theory} We introduce the tensor-like fields~$\tensor{e}{^i_\mu}$ and~$\tensor{e}{_i^\mu}$ as the components of the co-tetrad and tetrad fields, respectively. These fields are associated with Roman Lorentz (i.e. anholonomic, not associated with any coordinates) indices, whereas the Greek indices used hitherto throughout this paper are holonomic, i.e. corresponding to spacetime coordinates. We recover the standard spacetime metric from these fields using~$\tensor{e}{^i_\mu}\tensor{e}{^j_\nu}\tensor{\eta}{_{ij}}\equiv\tensor{g}{_{\mu\nu}}$ and the inverse~$\tensor{e}{_i^\mu}\tensor{e}{_j^\nu}\tensor{\eta}{^{ij}}\equiv\tensor{g}{^{\mu\nu}}$ with two identities~$\tensor{e}{^i_\mu}\tensor{e}{_i^\nu}\equiv\tensor*{\delta}{_\mu^\nu}$ and~$\tensor{e}{^i_\mu}\tensor{e}{_j^\mu}\equiv\tensor*{\delta}{_j^i}$ as extra kinematical restrictions on the components. We next introduce an independent spin connection field~$\tensor{\omega}{^{ij}_\mu}\equiv\tensor{\omega}{^{[ij]}_\mu}$. Equipped with these two sets of gauge fields, we can proceed to define the PGT torsion and PGT curvature as covariant field strengths
\begin{subequations}
\begin{align}
\ECT{^k_{ij}}&\equiv 2\tensor{e}{_i^\mu}\tensor{e}{_j^\nu}\big(\PD{_{[\mu|}}\tensor{e}{^k_{|\nu]}}+\tensor{\omega}{^k_{m[\mu|}}\tensor{e}{^m_{|\nu]}}\big),\label{PGTTorsion}
\\
 \tensor{\mathcal{R}}{^{kl}_{ij}}&\equiv 2\tensor{e}{_i^\mu}\tensor{e}{_j^\nu}\big(\PD{_{[\mu|}}\tensor{\omega}{^{kl}_{|\nu]}}+\tensor{\omega}{^k_{m[\mu|}}\tensor{\omega}{^{ml}_{|\nu]}}\big).\label{PGTCurvature}
\end{align}
\end{subequations}
Here, covariance refers to the transformations of both fields under the action of the Poincar\'e group, which we do not show in detail. In brief, however, under spatial translations the Greek indices transform with the standard Jacobian matrices of coordinate diffeomorphisms, and under local Lorentz transformations the Roman indices are rotated by the Lorentz transformation matrix familiar from special relativity, whilst the spin connection undergoes an additional inhomogeneous transformation involving the four-gradient of this local Lorentz rotation matrix. The holonomic components of the torsion and curvature in~\cref{eq:TorCLagrangian} are related to the quantities in~\cref{PGTTorsion,PGTCurvature} through~$\ECT{^\alpha_{\mu}_{\nu}}\equiv\tensor{e}{^i_\mu}\tensor{e}{_k^\alpha}\tensor{e}{^j_\nu}\ECT{^k_{ij}}$ and~$\tensor{\mathcal{R}}{^\rho_\sigma_{\mu\nu}}\equiv\tensor{e}{^i_\mu}\tensor{e}{^j_\nu}\tensor{e}{_k^\rho}\tensor{e}{_l_\sigma}\tensor{\mathcal{R}}{^{k}_{lij}}$. In the weak-field regime near to Minkowski spacetime we take~$\tensor{\omega}{^{ij}_\mu}$ to be perturbatively small, and we define the \emph{exact} perturbation of the tetrad to be~$\tensor{e}{_i^\mu}\equiv\tensor*{\delta}{_i^\mu}+\tensor{f}{_i^\mu}$ --- note that this corresponds to the so-called `Kronecker' gauge choice of the Minkowski vacuum. It is easy to confirm that, at lowest order in the perturbations, the Greek and Roman indices become interchangeable. When we count components, we find that there are in total 16 degrees of freedom (d.o.f) in~$\tensor{f}{_i^\mu}$ and a further 24 d.o.f in~$\tensor{\omega}{^{ij}_\mu}$. As source currents conjugate to the tetrad perturbation~$\tensor{f}{_i^\mu}$ and spin connection~$\tensor{\omega}{^{ij}_\mu}$ we introduce the (asymmetric) stress-energy tensor of matter~$\tensor{\tau}{^i_\mu}$ and also the spin current of matter~$\tensor{\sigma}{_{ij}^\mu}$. These source currents have their origin in the matter Lagrangian, the form of which need not be specified.

\paragraph*{Post-Riemannian formulation} Whilst the gauge-theoretic structure of PGT is the motivating framework for the TorC model, it may not be the most familiar framework to the reader. Indeed, there is a conceptually far simpler way to work with the TorC model, whereby the theory is separated into those parts which depend on the same Riemannian geometry in which GR is formulated, and the \emph{post-Riemannian} parts which depend on the torsion. To make the comparison with textbook GR as seamless as possible, it is usual to formulate this \emph{post-Riemannian decomposition} in terms of the metric~$\tensor{g}{_{\mu\nu}}$ rather than the tetrad~$\tensor{e}{_i^\mu}$. The non-Riemannian (torsionful) curvature defined in~\cref{PGTCurvature} and illustrated in~\cref{fig:NonRiemannianSchematic} can be alternatively expressed in terms of the usual metric and coordinate indices, along with a torsionful affine connection~$\AffineConnection{^\rho_{\mu\nu}}$ as
\begin{align}
\tensor{\mathcal{R}}{^\rho_\sigma_{\mu\nu}} &\equiv 2\left(\tensor{\partial}{_{[\mu|}}\AffineConnection{^\rho_{|\nu]}_\sigma}+\AffineConnection{^\rho_{[\mu|}_\alpha}\AffineConnection{^\alpha_{|\nu]}_\sigma}\right) \,,\label{PostRiemannianCurvatureDef}
\end{align}
where the torsion defined in~\cref{PGTTorsion} and illustrated in~\cref{fig:NonRiemannianSchematic} may be alternatively defined via the affine connection itself as
\begin{equation}
	\ECT{^\alpha_\mu_\nu} \equiv 2\AffineConnection{^\alpha_{[\mu\nu]}} \,. \label{PostRiemannianTorsionDef}
\end{equation}
Note that~\cref{PostRiemannianCurvatureDef,PostRiemannianTorsionDef} are definitions that one is more likely to find in introductory textbooks. The physical content of~\cref{eq:TorCLagrangian} does not change when one uses~\cref{PostRiemannianCurvatureDef,PostRiemannianTorsionDef} instead of~\cref{PGTCurvature,PGTTorsion}, so long as~$\tensor{g}{_{\mu\nu}}$ and~$\AffineConnection{^\mu_{\nu\sigma}}$ are used as fundamental fields instead of~$\tensor{e}{_i^\mu}$ and~$\tensor{\omega}{^{ij}_\mu}$, respectively. The use of the affine connection as a fundamental field is known as the \emph{Palatini} formulation of the theory. The Palatini approach is occasionally used for GR itself, but this is not the standard approach, and once again we seek to connect with the most standard formulation possible. To reach this final formulation, one can decompose the affine connection into a Riemannian part and a post-Riemannian (torsionful) part
\begin{equation}
\AffineConnection{^\rho_\mu_\nu} \equiv \LeviCivitaConnection{^\rho_\mu_\nu} + \frac{1}{2}\left(\ECT{^\rho_{\mu\nu}} - \ECT{_{\nu\mu}^\rho} + \ECT{_{\mu}^\rho_{\nu}}\right),\label{Contorsion}
\end{equation}
where~$\LeviCivitaConnection{^\nu_\mu_\rho}\equiv\tensor{g}{^{\nu\lambda}}\big(\PD{_{(\mu}}\tensor{g}{_{\rho)\lambda}}-\frac{1}{2}\PD{_{\lambda}}\tensor{g}{_{\mu\rho}}\big)$ is the familiar Levi--Civita connection, i.e. the usual Christoffel symbols. It is easy to check that~\cref{Contorsion,PostRiemannianTorsionDef} are mutually consistent. When~\cref{Contorsion} is plugged into~\cref{PostRiemannianCurvatureDef}, one obtains a long formula (the details of whose indices are not useful to our discussion) of the form
\begin{equation}\label{PRE}
	\tensor{\mathcal{R}}{}\sim\RiemannianR{}+\ECT{}^2+\CD{}\ECT{},
\end{equation}
where~$\CD{}$ denotes the usual GR covariant derivative with connection~$\LeviCivitaConnection{^\nu_\mu_\rho}$ and where the Riemannian curvature tensor of GR is
\begin{align}
\RiemannianR{^\rho_\sigma_{\mu\nu}} &\equiv 2\left(\tensor{\partial}{_{[\mu|}}\LeviCivitaConnection{^\rho_{|\nu]}_\sigma}+\LeviCivitaConnection{^\rho_{[\mu|}_\alpha}\LeviCivitaConnection{^\alpha_{|\nu]}_\sigma}\right) \,.\label{RiemannianCurvatureDef}
\end{align}
One may plug the post-Riemannian expansion in~\cref{PRE} into~\cref{eq:TorCLagrangian} to obtain the TorC Lagrangian in its post-Riemannian form. 

\paragraph*{Torsion condensation} The vastness of the resulting post-Riemannian expression, which contains very many terms but which takes the schematic form
\begin{equation}\label{PRELagrangian}
	\begin{aligned}
		\mathcal{L}_{\text{TorC}}&\sim
		\mu\RiemannianR{}^2
		+\mu\RiemannianR{}\ECT{}^2
		+\mu\RiemannianR{}\CD{}\ECT{}
				 \\&\quad
		+\mu\left(\CD{}\ECT{}\right)^2
		+(\mu+\nu)\ECT{}^2\CD{}\ECT{}
				 \\&\quad
		+\left(\Mp+\mu\lambda\right)\ECT{}^2
		+(\mu+\nu)\ECT{}^4
				 \\&\quad
		- \Mp \Lambda + \mathcal{L}_{\mathrm{M}}
		,
	\end{aligned}
\end{equation}
reflects the true non-linear richness of the TorC model as expressed its easiest-to-understand field variables~$\tensor{g}{_{\mu\nu}}$ and~$\ECT{^\mu_{\nu\sigma}}$. In essence, we see how TorC manifests as a gravity model in which the usual Riemannian curvature tensor is coupled in a highly non-minimal way to~$\ECT{^\mu_{\nu\sigma}}$, as if the latter were an exotic kind of matter field. One takes variations of~\cref{PRELagrangian} with respect to~$\tensor{g}{_{\mu\nu}}$ to obtain the (highly modified) Einstein field equations of TorC; in particular the metric variation of~$\mathcal{L}_{\mathrm{M}}$ leads to the stress-energy tensor~$\StressEnergyTensor{^{\mu\nu}}$ of actual matter (baryons and CDM). The variations of~\cref{PRELagrangian} with respect to~$\ECT{^\mu_{\nu\sigma}}$ lead to the torsion equations of motion; note that the torsion variation of~$\mathcal{L}_{\mathrm{M}}$ will be denoted presently by the current~$\TorsionSource{_\mu^{\nu\sigma}}$ --- unlike the matter spin tensor conjugate to the spin connection, and the so-called hypermomentum current conjugate to the affine connection, this torsion source current does not have a standard name in the literature. Regardless of this source current and its origins in the matter sector, the torsion dynamics in the gravity sector clearly tell us that torsion tends to condense at the vacuum expectation value~$\TT^2 \sim \Mp$. To see this easily from~\cref{PRELagrangian} note that the torsion has both a quadratic interaction and a quartic interaction. This feature is shared by the `sombrero' potential of the Higgs doublet, where it leads to electroweak symmetry breaking.\footnote{Note that we assume the separation of scales~$\Mp\gg\lambda$.} Heuristically, one may see how the Einstein--Hilbert-like limit emerges from infrared physics when this condensate is plugged back into~\cref{PRELagrangian}, by examining the second term, with the quadratic-curvature operators in the first term acting as short-distance corrections.

\begin{figure*}[htbp]
	\includegraphics[width=\linewidth]{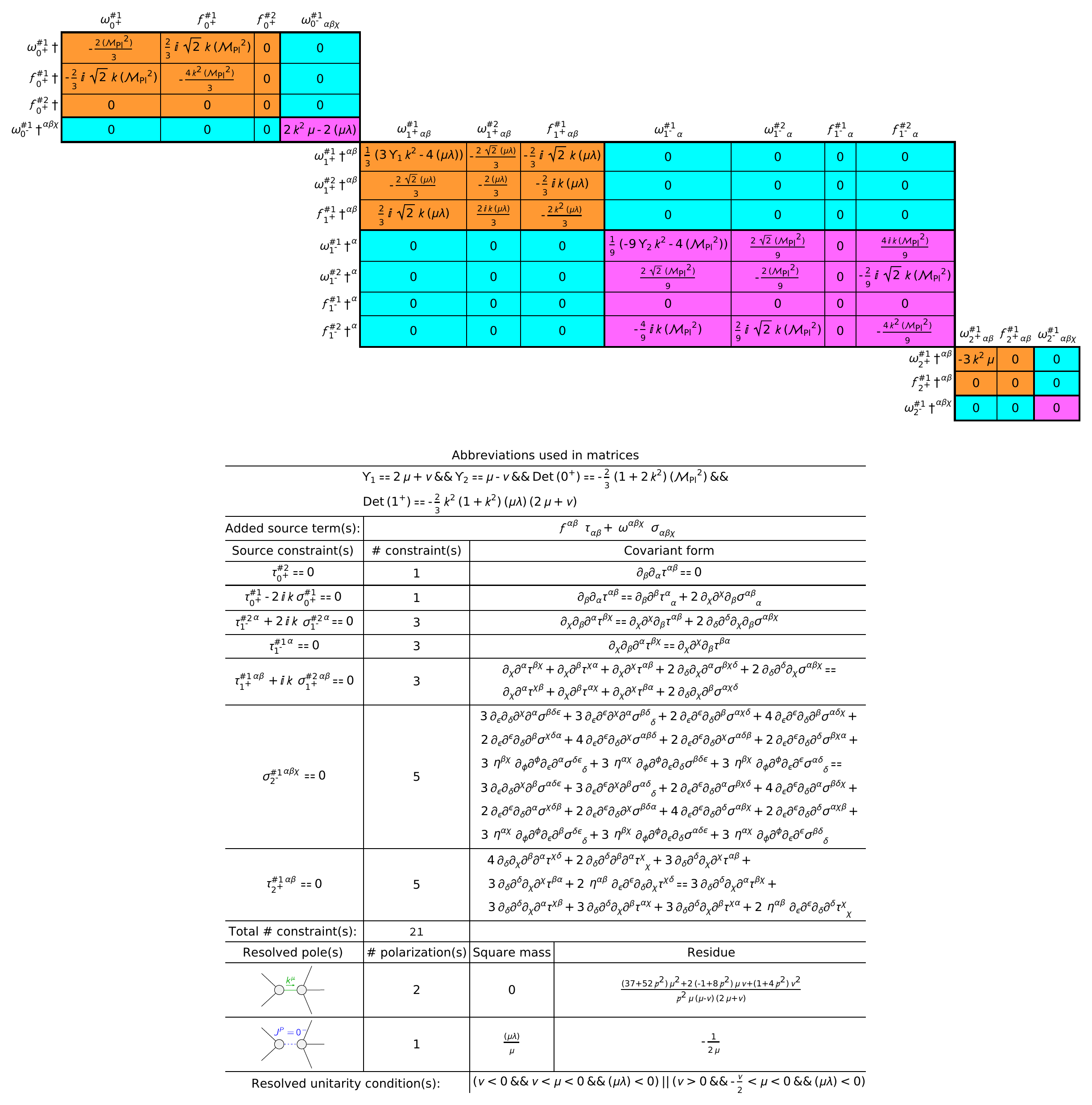}
	\caption{Output generated by \textit{PSALTer}. The spectrograph of the theory defined by~\cref{eq:TorCLagrangian}. At the top, the (Fourier transformed) wave operator~$\mathsf{O}(k)$ from~\cref{GenLag} is displayed in the~$J^P$-representation defined in~\cref{FieldKinematicsTetradPerturbation,FieldKinematicsSpinConnection}. The~$J^P$ states are organised into diagonal blocks of~$J=0$,~$J=1$ and~$J=2$, and within each block into diagonal parity-preserving sectors of~$P=1$ (orange) and~$P=-1$ (purple), along with off-diagonal parity-violating blocks (teal). The Lagrangian in~\cref{eq:TorCLagrangian} does not contain any odd powers of the totally antisymmetric tensor, so there is no mixing between parities in this theory. The theory contains as many as 21 gauge generators: this far exceeds the 10 gauge generators common to all instances of Poincar\'e gauge theory. The gauge symmetries impose constraints on the source currents. The particle spectrum consists of two massless polarisations and one massive pseudoscalar particle (i.e.~$J^P=0^-$). In terms of the couplings in~\cref{eq:TorCLagrangian} the pseudoscalar has mass~$\sqrt{\lambda}$, so that it is not a tachyon if~$\lambda\geq 0$. Its pole residue indicates that it is not a ghost if~$\mu<0$. The residue of the massless pole (where~$p$ denotes the particle energy in some fiducial frame) then indicates a no-ghost condition of~$(\nu + 2 \mu)(\nu - \mu) > 0$. Together, these three unitarity conditions are consistent with those in~\cref{eq:TorCconditions}.}
\label{ParticleSpectrographTorCPGT}
\end{figure*}

\begin{figure*}[htbp]
	\includegraphics[width=0.9\linewidth]{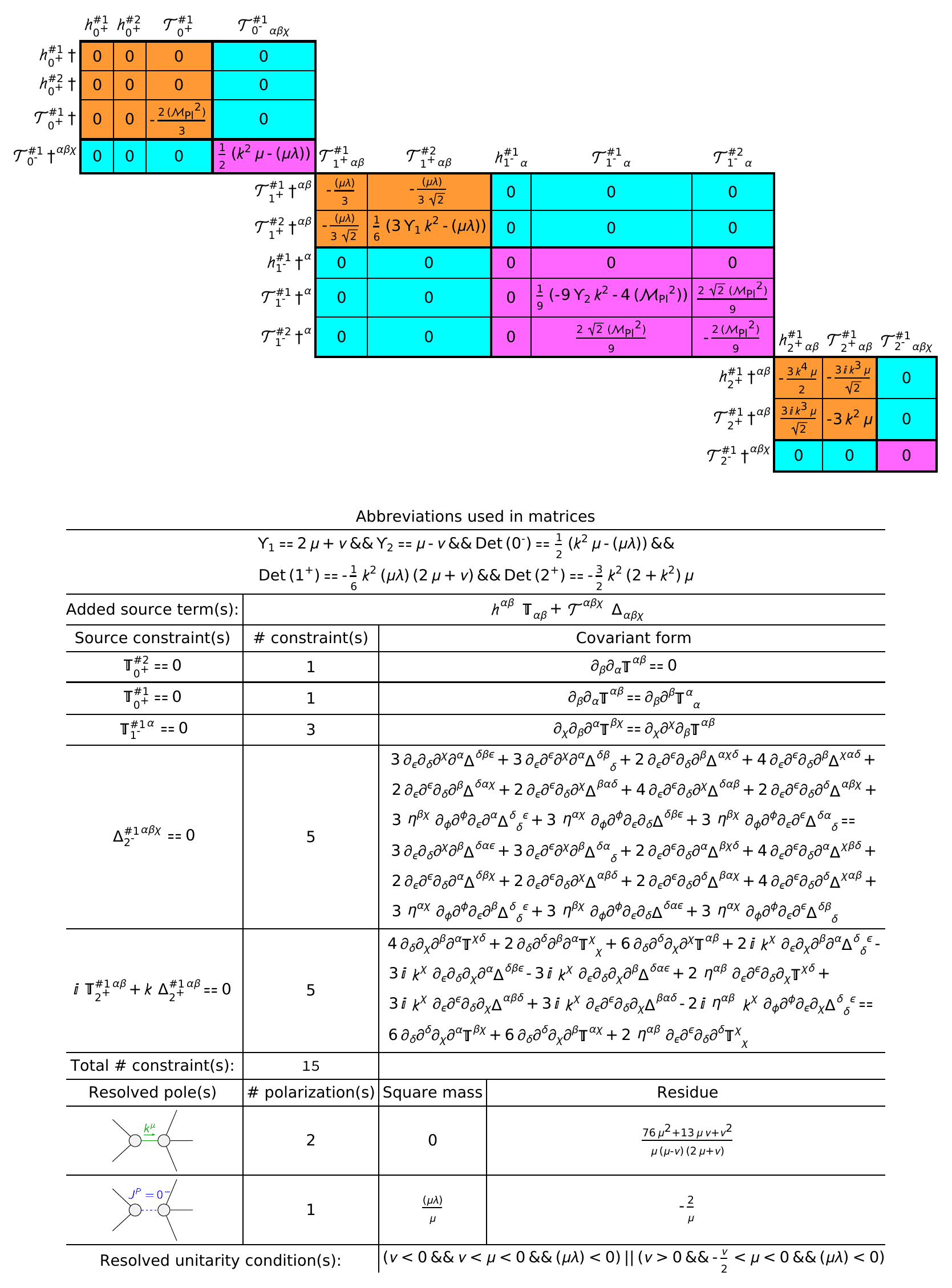}
	\caption{Output generated by \textit{PSALTer}. The spectrograph of the theory defined by~\cref{PRELagrangian}, i.e. the post-Riemannian formulation of TorC. Due to the elimination of redundant (gauge) field d.o.f in~\cref{FieldKinematicsMetricPerturbation}, the wave operator matrix is smaller than in~\cref{ParticleSpectrographTorCPGT}. To compensate, there are six \emph{fewer} gauge symmetries and corresponding source constraints (15 symmetry generators rather than 21, though this still exceeds the four expected diffeomeorphism generators due to the presence of accidental symmetries in the linearised theory). Despite these changes, the physical content of TorC is unchanged, namely the (measurable) pseudoscalar mass, the number of massless polarisations and the overall unitarity conditions. These conditions are consistent with those in~\cref{eq:TorCconditions,ParticleSpectrographTorCPGT}. See~\cref{FieldKinematicsMetricPerturbation,FieldKinematicsTPerturbation} for the definitions of the fields and sources used in this spectrograph.}
\label{ParticleSpectrographTorCECT}
\end{figure*}

\paragraph*{Particle spectroscopy} The spectrum of gravitational perturbations exhibited by TorC around the torsion condensate is an area of active study, well beyond the scope of this initial work; this naturally includes the development of the associated theory of cosmological perturbations. The tree-level particle spectrum is, however, very well understood around Minkowski spacetime with \emph{vanishing} background torsion: it is the unitarity of this particle spectrum which single-handedly motivates parameter constraints in~\cref{eq:TorCconditions}, and so in this section we improve on~\cite{Lin:2019ugq} by re-deriving these constraints using modern methods.\footnote{Note that even these constraints may possibly be relaxed if TorC is proven to be inherently non-perturbative around the Minkowski background, a contingency which (unlike for standard, non-condensate models) would not jeapordise its phenomenological utility.} In order to determine the tree-level spectrum of the theory in~\cref{eq:TorCLagrangian}, we use the \textit{PSALTer} software~\cite{Barker:2024juc,Barker:2025qmw}. Let us start with the formulation of TorC as a PGT. A collection of completely general test source currents~$\tensor{\sigma}{_{ij}^\mu}$ and~$\tensor{\tau}{^i_\mu}$ are together introduced as the vector~$\mathsf{J}$, whilst the fields~$\tensor{f}{_i^\mu}$ and~$\tensor{\omega}{^{ij}_\mu}$ are denoted by another vector~$\zeta$, so that the whole matter Lagrangian is simply~$\mathcal{L}_{\text{M}}\sim\zeta^{\text{T}}\cdot\mathsf{J}$ when working to lowest order in perturbations. Basing ourselves in position space, it follows that the action corresponding to~\cref{eq:TorCLagrangian} can be expressed schematically in terms of these source and field vectors as
\begin{equation}\label{GenLag}
	\mathcal{S}=\int\mathrm{d}^4x\ \zeta^{\text{T}}(x)\cdot\left[\mathsf{O}(\partial)\cdot\zeta(x)-\mathsf{J}(x)\right]+\mathcal{O}(\zeta^3),
\end{equation}
where the operator matrix~$\mathsf{O}(\partial)$ is termed the (linearized) \emph{wave operator}. This matrix corresponds to a tensor structure which is a polynomial of (up to) second order in the partial derivative~$\PD{_\mu}$. It is also parameterised by~$\mu$ and~$\nu$, along with~$\lambda$ and~$\Mp$. Working from this representation of the action, the saturated propagator derived from~\cref{GenLag} is extracted by moving over to momentum space via~$\PD{_\mu}\mapsto -i\tensor{k}{_\mu}$, before proceeding to `invert' the wave operator and sandwich it between the conserved sources according to the formula
\begin{equation}\label{Propagator}
	\Pi(k)\equiv\mathsf{J}^\dagger(k)\cdot\mathsf{O}^{-1}(k)\cdot\mathsf{J}(k).
\end{equation}
As expected of a PGT (and especially true in the case of TorC, for which extra `accidental' symmetries appear in the linearisation) the model in~\cref{GenLag} turns out to feature various gauge invariances: this means that the supposed `inverse' in~\cref{Propagator} is not well defined owing to degeneracies in the matrix representation of the wave operator. As a general principle, the same gauge symmetries that are associated one-to-one with these degeneracies result in the presence of constraints on the physical values of the source currents. Thus, the consequence of the `sandwiching' procedure in~\cref{Propagator} is that the singular parts of the matrix inverse exactly cancel against the roots of the sources. Continuing to work in~$k$-space, standard methods in polology tell us how the locations of the poles in~$\Pi(k)$ describe the (squares of the) masses of eternal states. Specifically, real masses are needed for non-tachyonic (physical) particles. At the same time, the residue of~$\Pi(k)$ on each of the poles must be positive-definite in order for the state to not be a ghost. In the case of the massive quantum states, use of a spin-parity ($J^P$) basis is ideal for framing the problem of wave operator inversion in terms of basic linear algebra. In the case of massless states, on the other hand, spin is properly described by helicity quantum numbers. For the massless calculation the spectrum is computed by brute-force, using the individual components of the tensors. Whilst this component-level procedure is effective at extracting the number of polarisations and their corresponding unitarity conditions, the massless~$J$ and~$P$ numbers are not always apparent. The~$J^P$ modes contained within~$\tensor{f}{_i^\mu}$ and~$\tensor{\omega}{^{ij}_\mu}$ are shown in~\cref{FieldKinematicsTetradPerturbation,FieldKinematicsSpinConnection} respectively. The spectrograph in~\cref{ParticleSpectrographTorCPGT} is then produced automatically by feeding in the linearisation of the TorC Lagrangian density in~\cref{eq:TorCLagrangian} to the software. One can also perform the corresponding analysis of the post-Riemannian decomposed model in~\cref{PRELagrangian}, and this provides a reassuring cross-check. In this case, one uses the (exactly defined) metric perturbation~$\MetricPerturbation{_{\mu\nu}}\equiv\tensor{g}{_{\mu\nu}}-\tensor{\eta}{_{\mu\nu}}$ and small~$\ECT{^\mu_{\nu\sigma}}$ instead of~$\tensor{f}{_i^\mu}$ and~$\tensor{\omega}{^{ij}_\mu}$, and the sources are~$\StressEnergyTensor{^{\mu\nu}}$ and~$\TorsionSource{_\mu^{\nu\sigma}}$. The spectrograph in this case is shown in~\cref{ParticleSpectrographTorCECT}. The kinematics tables for the fields and their sources are shown in~\cref{FieldKinematicsMetricPerturbation,FieldKinematicsTPerturbation}. The sources used for this spectroscopy analysis can be found at~\cite{PSALTer}.

\begin{table*}[htbp]
	\includegraphics[width=\linewidth]{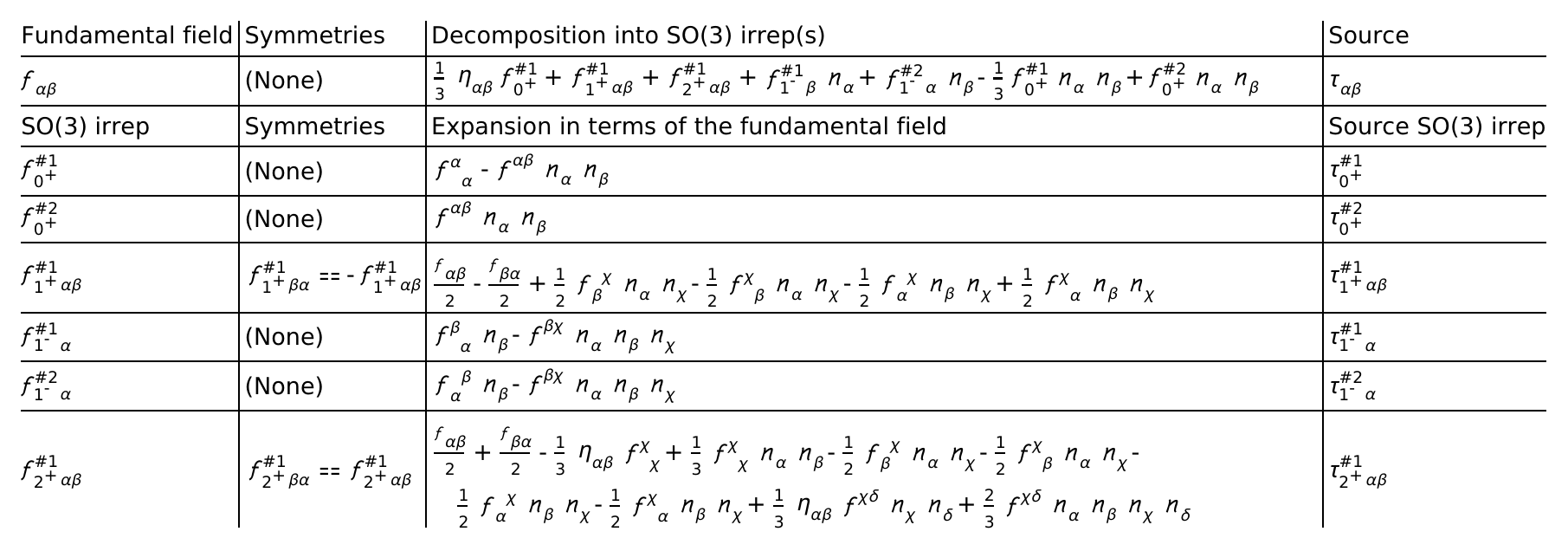}
	\caption{Output generated by \textit{PSALTer}. Field kinematics of the tetrad perturbation~$\tensor{f}{_i^\nu}$ and its conjugate source current~$\tensor{\tau}{^i_\mu}$ --- physically interpreted as the asymmetric stress-energy tensor of the matter sector. Note that when working to lowest order in the perturbations, the Greek (coordinate) and Roman (Lorentz) indices are freely exchanged. The tetrad perturbation decomposes into irreducible parts under the action of the three-dimensional rotation group~$\mathrm{SO}(3)$. These correspond to states of definite spin~$J$ and parity~$P$ (collectively~$J^P$). This decomposition is made possible by the introduction of a preferred frame, i.e. a unit-timelike vector~$\tensor{n}{_{\mu}}\equiv\tensor{k}{_{\mu}}/\sqrt{\tensor{k}{^\nu}\tensor{k}{_\nu}}$, where~$\tensor{k}{_\mu}$ is the four-momentum of massive particles. These definitions are used in~\cref{ParticleSpectrographTorCPGT}.}
	\label{FieldKinematicsTetradPerturbation}
\end{table*}

\begin{table*}[htbp]
	\includegraphics[width=\linewidth]{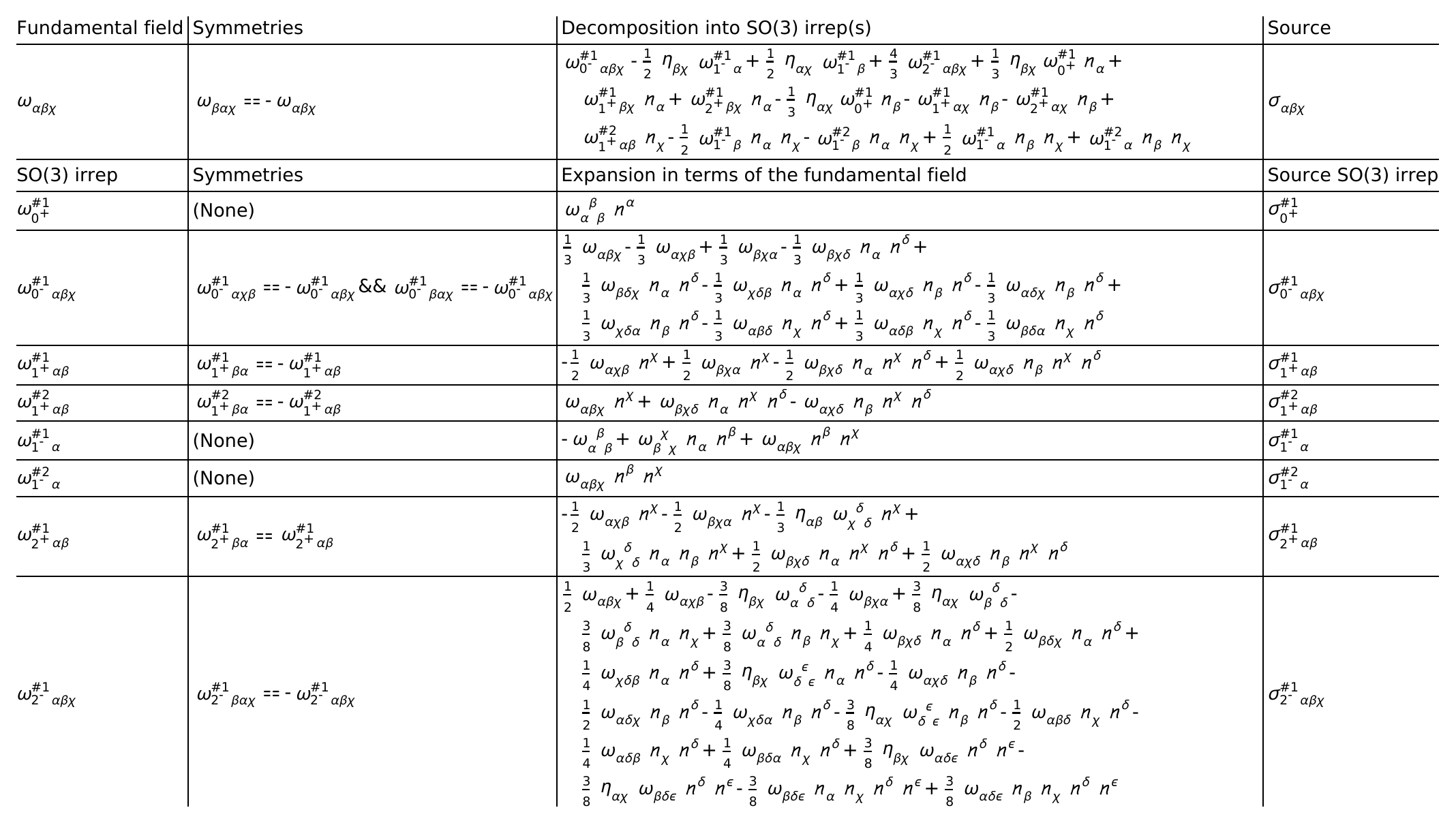}
	\caption{Output generated by \textit{PSALTer}. Field kinematics of the spin connection~$\tensor{\omega}{^{ij}_\mu}\equiv\tensor{\omega}{^{[ij]}_\mu}$ and conjugate source~$\tensor{\sigma}{_{ij}^\mu}$ --- the spin current of matter. The principles behind the decomposition are precisely the same as in~\cref{FieldKinematicsTetradPerturbation}. These definitions are used in~\cref{ParticleSpectrographTorCPGT}.}
	\label{FieldKinematicsSpinConnection}
\end{table*} 

\begin{table*}[htbp]
	\includegraphics[width=\linewidth]{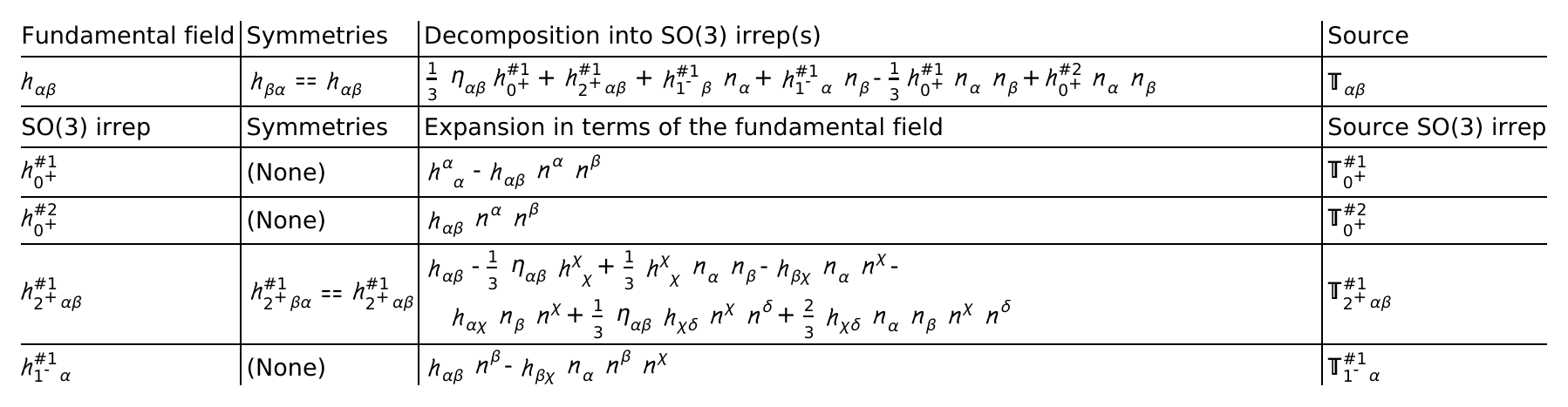}
	\caption{Output generated by \textit{PSALTer}. Field kinematics of the metric perturbation~$\MetricPerturbation{_{\mu\nu}}$ and conjugate source~$\StressEnergyTensor{^{\mu\nu}}$. Note that, compared to~\cref{FieldKinematicsTetradPerturbation}, the metric perturbation has six \emph{fewer} d.o.f, manifest as the loss of one~$1^+$ and one~$1^-$ modes, and owing to the fact that the tetrad is an asymmetric rank-two field, whilst the metric is symmetric. How is it possible that the physics is not lost by this reduction of the fundamental fields? The post-Riemannian metric formulation of TorC is, in some sense, a description of the PGT formulation of TorC in gauge-invariant variables. In these variables, the local symmetry of the theory under the Lorentz part of the Poincar\'e group becomes hidden, so that the only symmetry expected of~\cref{PRELagrangian} is diffeomorphism invariance as shared by GR (encoding the translational part of the Poincar\'e group). In other words, the metric does away with redundant (gauge) d.o.f in the tetrad. These definitions are used in~\cref{ParticleSpectrographTorCECT}.}
	\label{FieldKinematicsMetricPerturbation}
\end{table*}

\begin{table*}[htbp]
	\includegraphics[width=\linewidth]{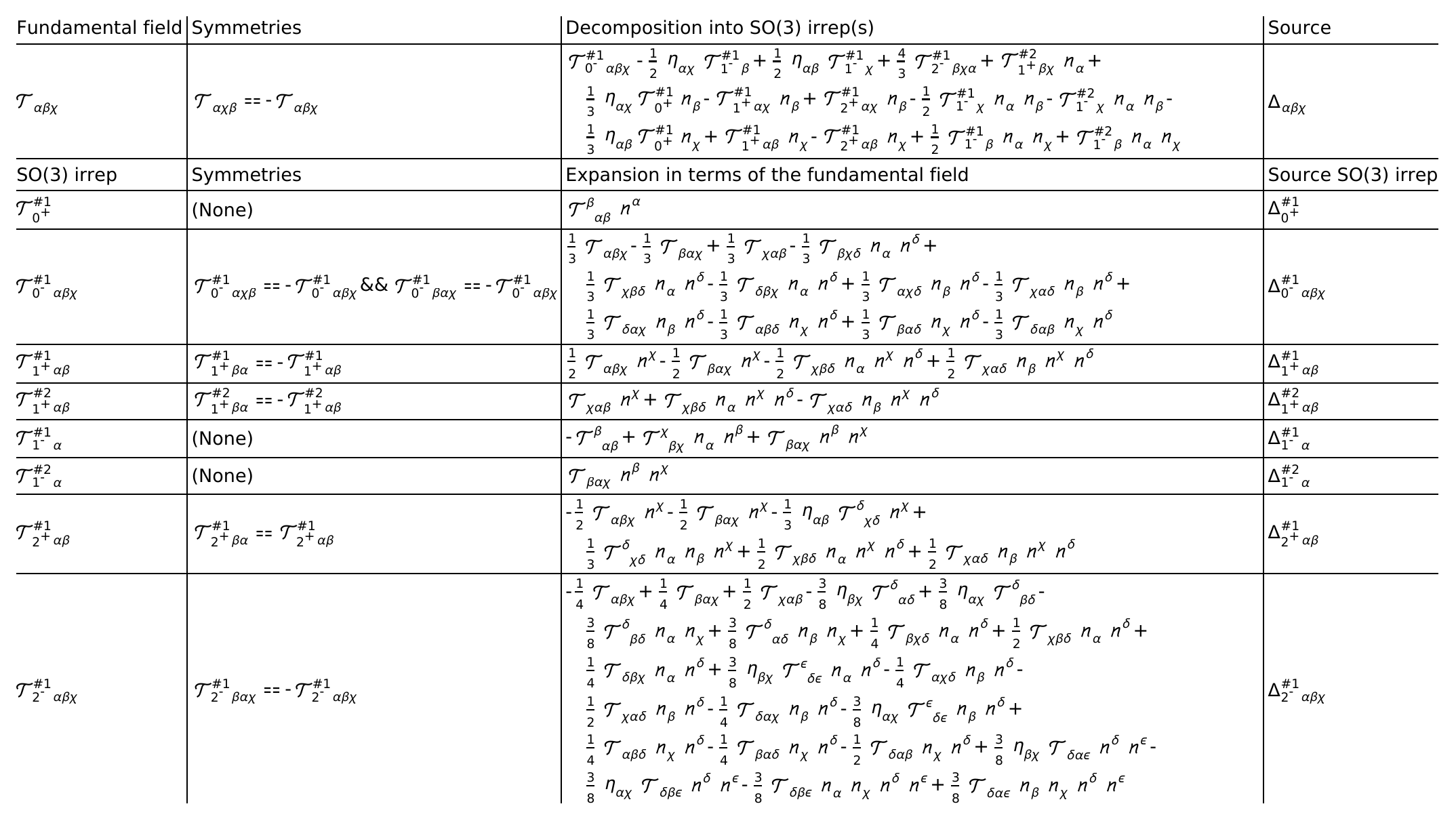}
	\caption{Output generated by \textit{PSALTer}. Field kinematics of the torsional perturbation~$\ECT{^\mu_{\nu\sigma}}$ and conjugate source~$\TorsionSource{_\mu^{\nu\sigma}}$. Note that since the torsion field~$\ECT{^\mu_{\nu\sigma}}\equiv\ECT{^\mu_{[\nu\sigma]}}$ is antisymmetric, it has 24 independent d.o.f, which is the same as the number of d.o.f in the spin connection~$\tensor{\omega}{^{ij}_\mu}\equiv\tensor{\omega}{^{[ij]}_\mu}$. Thus, the~$J^P$ modes are basically `clones' of those in~\cref{FieldKinematicsSpinConnection}. These definitions are used in~\cref{ParticleSpectrographTorCECT}.}
	\label{FieldKinematicsTPerturbation}
\end{table*} 

\noindent\begin{minipage}[t]{0.48\textwidth}
\section{Initial conditions}\label{sec:TorCpower}
By writing~$a$ and~$\varpi$ as~$a = \sum_{i=1}^{n} C_i^a \tau^i$ and~$\varpi = \sum_{i=1}^{n} C_i^{\varpi} \tau^i$, one can substitute the power series ans\"atze into the TorC field equations to determine their coefficients, thereby fully specifying the early-time behaviour of both~$a$ and~$\varpi$. The solutions for the coefficients of the first few terms in these series are:
\begin{samepage}
\begin{subequations}
	\begin{align}
			a = & \frac{H_0 \sqrt{\Omega_\mathrm{r}}}{\alpha \varpi_\mathrm{r}}\tau + \frac{H_0^2 \Omega_\mathrm{m}\bigl(3 \varpi_\mathrm{r}^2 + 1\bigl)}{16 \alpha \varpi_\mathrm{r}^2} \tau^2 \nonumber\\
				& + \frac{5 H_0^3 \Omega_\mathrm{m}^2 \bigl(\varpi_\mathrm{r}^2-1\bigr)}{512 \alpha \varpi_\mathrm{r}^3\sqrt{\Omega_\mathrm{r}}} \tau^3 \nonumber\\
				& + \frac{H_0^4 \bigl(\varpi_\mathrm{r}^2-1\bigr) \bigl(27 \varpi_\mathrm{r}^2 -121\bigr) \Omega_\mathrm{m}^3}{49152 \alpha \varpi_\mathrm{r}^4\Omega_\mathrm{r}} \tau^4 \nonumber\\
				& + \frac{H_0^5}{131720 \alpha \varpi_\mathrm{r}^5 \Omega_\mathrm{r}^{3/2}} \biggl(32768 \bigl(3 \varpi_\mathrm{r}^2 + 1\bigr) \Omega_\mathrm{r}^3 \Omega_\Lambda \nonumber\\
				& - 49 \bigl(9 \varpi_\mathrm{r}^4 - 29 \varpi_\mathrm{r}^2 + 20 \bigr) \Omega_\mathrm{m}^4 \biggr) \tau^5 + \mathcal{O}(\tau^6),
		 \label{eq:aPowerSeries}\\
					\varpi = & \varpi_\mathrm{r} - \frac{3 H_0 \bigl(\varpi_\mathrm{r}^2-1\bigr)\Omega_\mathrm{m}}{16 \sqrt{\Omega_\mathrm{r}}} \tau \nonumber\\
							 & + \frac{H_0^2 \bigl(18 \varpi_\mathrm{r}^4 - 5 \varpi_\mathrm{r}^2 - 13 \bigr) \Omega_\mathrm{m}^2}{512 \varpi_\mathrm{r} \Omega_\mathrm{r}} \tau^2 \nonumber\\
							 & + \frac{H_0^3 \bigl(-324 \varpi_\mathrm{r}^6 + 45 \varpi_\mathrm{r}^4 - 20 \varpi_\mathrm{r}^2 + 299 \bigr) \Omega_\mathrm{m}^3}{49152 \varpi_\mathrm{r}^2 \Omega_\mathrm{r}^{3/2}} \tau^3 \nonumber\\
							 & + \frac{H_0^4}{1310720 \varpi_\mathrm{r}^3 \Omega_\mathrm{r}^2} \biggl[- 98304 \Omega_\mathrm{r}^3 \Omega_\Lambda \nonumber\\
							 & +\bigl(\varpi_\mathrm{r}^2 -1\bigr) \biggl(\bigl(2 \varpi_\mathrm{r}^2\bigl(810 (\varpi_\mathrm{r}^2 + \varpi_\mathrm{r}^4) + 731\bigr)\nonumber \\
					& + 2327\bigr)\biggr) \Omega_\mathrm{m}^4 \biggr] \tau^4 + \mathcal{O}(\tau^5).
		 \label{eq:varpiPowerSeries}
	\end{align}
\end{subequations}
\end{samepage}
The parameter~$\varpi_\mathrm{r}$ corresponds to the zeroth-order term in the power series of~$\varpi$. As such, the power series of~$a$ and~$\varpi$ are functions of~$\varpi_\mathrm{r}$ and~$\tau$.
Note that the normalisation~$\alpha$ only affects the scale factor~$a$. To ensure the validity of the power series, the initial value of~$\tau$ used for numerical evolution is chosen to be sufficiently small such that the first term in the power series of~$a$ dominates over the second term
\begin{equation}
		\tau_{\mathrm{ini}} = 10^{-6} \frac{16 \sqrt{\Omega_\mathrm{r}} \varpi_\mathrm{r}}{\Omega_\mathrm{m} \bigl(3 \varpi_\mathrm{r}^2 + 1\bigr) H_0}.
\end{equation}
At such an early time, it is clear that the system's evolution is primarily governed by the zeroth-order term in the power series expansion of the torsion scalar field,~$\varpi_\mathrm{r}$.
\end{minipage}%
\hfill%

\section{A discussion on Big Bang nucleosynthesis}\label{sec:Neff}
As discussed in~\cite{Barker:2020gcp} and illustrated in~\cref{fig:densityevolution}, the presence of torsion via the parameter $\varpi_r$ in TorC cosmology manifests as an early dark radiation component. This contribution to the early universe expansion rate can be parameterised in terms of the effective number of neutrino species,
\begin{equation}
		N_\mathrm{eff} = N_\nu + \Delta N_\mathrm{eff}.
\end{equation}
Here,~$N_\nu = 3.046$~\cite{Mangano:2005cc} accounts for the standard contribution from the three neutrino species, including the corrections from non-instantaneous neutrino decoupling and flavour oscillations. The specific contribution from torsion is given by~\cite{Barker:2020gcp}
\begin{equation}
		\Delta N_\mathrm{\varpi_r,eff} = \left(\varpi_r^{-2}-1\right)\left(\frac{8}{7}\left(\frac{11}{4}\right)^{4/3}+N_\mathrm{eff}\right).
\end{equation}
Big Bang nucleosynthesis (BBN), the process responsible for the formation of light elements, depends sensitively on the nuclear reaction rates and the Hubble expansion rate at early radiation dominated epochs. An increase in the expansion rate leads to an earlier freeze-out of weak interactions, resulting in a higher neutron-to-proton ratio and consequently a higher abundance of helium-4 and deuterium~\cite{Steigman:2007xt,Cyburt:2015mya}. Therefore, a modification to~$N_\mathrm{eff}$ due to torsion would impact the predicted primordial abundances of light elements.  The primordial helium abundance, $Y_\mathrm{P}$, subsequently influences the damping tail of the CMB power spectra~\cite{Cooke:2024nqz}. 

It is important to note that directly modifying the input value of $N_\mathrm{eff}$ in our Boltzmann solver, while simultaneously parameterising TorC as a modified dark energy component, would result in double counting the effect of torsion on the background expansion.
Therefore, in the main analysis, the derived~$\Delta N_\mathrm{eff}$ is excluded in the cosmological evolution. However, to verify whether the torsion-induced shift in primordial abundances impacts our conclusions, we performed a validation run. In this run, the helium mass fraction, $Y_\mathrm{P}$, was derived using the calculated~$\Delta N_\mathrm{eff}$ through BBN predictions from \texttt{PArthENoPE}~\cite{Pisanti:2007hk,Consiglio:2017pot} as implemented in \texttt{CAMB}. This approach accounts for the torsion-induced shift in~$N_\mathrm{eff}$ solely through its impact on primordial abundances. It is found that the inclusion of this effect does not significantly alter the main conclusions of this paper.

\section{Density evolution varying~$\Omega_\Lambda$}\label{sec:matter-evolution}
The evolution of the density parameters as a function of~$\Omega_\Lambda$ is shown in~\cref{fig:matter-evolution-OmL}. The top panel corresponds to the standard \LCDM{} case, where the dark energy density is fixed by the densities of the other components at~$a=1$. The middle panel shows the TorC model with~$\varpi_\mathrm{r}$ held at its \LCDM{}-correspondent value, and with a higher choice of~$\Omega_\Lambda = 1.3$ to illustrate its impact on the density evolution. In TorC, this is implemented through a renormalisation of the scale factor~$a$, as described in~\cref{sec:TorCparams} and~\cref{eq:rescale}. Numerically, the system is evolved until the Hubble parameter matches the specified value of~$H_0$, the corresponding scale factor is identified, and the scale factor is then rescaled so that $H = H_0$ at $a = 1$. This procedure allows~$\Omega_\Lambda$ to be treated as an independent parameter in the TorC model. Varying~$\Omega_\Lambda$ shifts both the matter–radiation and matter–dark energy equality epochs. For larger~$\Omega_\Lambda$, matter–dark energy equality occurs earlier, while matter–radiation equality is pushed later, compressing the matter-dominated era. The bottom panel shows the effective equation of state of the modified dark energy component: it behaves like matter ($w\simeq 0$) at early times and transitions to a dark-energy-like phase ($w\simeq -1$) at late times. This transition introduces a pole in the equation of state, motivating the use of density and pressure variables in the main analysis.
\begin{center}
	\includegraphics[width=\linewidth]{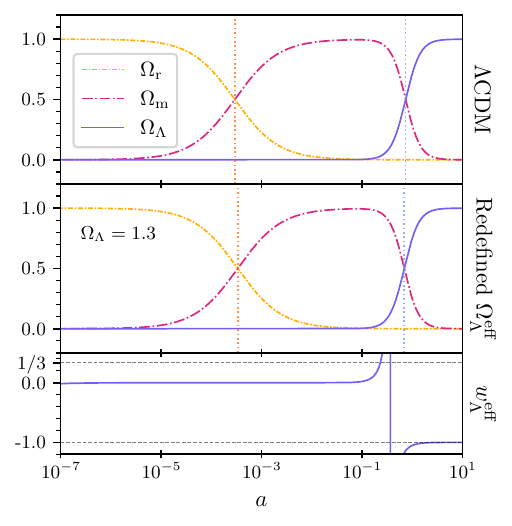}
	\captionof{figure}{Evolution of the density parameters and the effective dark-energy equation of state in \LCDM{} and TorC cosmologies for varying~$\Omega_\Lambda$. The top panel shows the standard \LCDM{} evolution with $\Omega_\Lambda$ determined from the remaining components at $a = 1$. The middle panel shows the TorC evolution with~$\varpi_\mathrm{r}$ fixed to its \LCDM{}-correspondence value and with an increased choice of~$\Omega_\Lambda = 1.3$. In these two panels, the density evolution of radiation (yellow), matter (red), and dark energy (purple) is shown. The bottom panel presents the effective equation of state of the modified dark-energy component in TorC, incorporating the rescaling of the scale factor~$a$. The other density parameters are held fixed to their Planck 2018 best-fit values~\cite{Planck:2018vyg}.}
	\label{fig:matter-evolution-OmL}
\end{center}

\section{Hubble parameter posteriors}\label{sec:hubble-evolution}
\begin{figure}[!htbp]
\centering
\includegraphics[width=\linewidth]{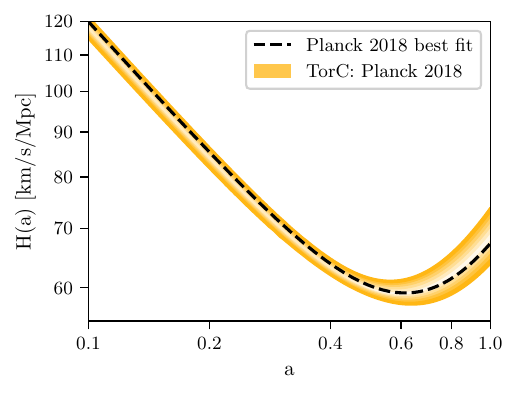}
\vspace{0.5cm}
\includegraphics[width=\linewidth]{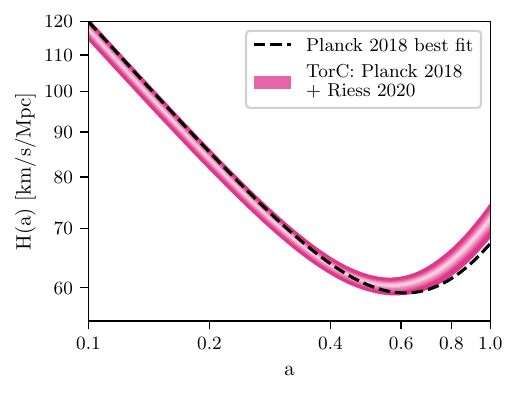}
\caption{Posterior evolution of the Hubble parameter,~$H(a)$, as a function of the scale factor~$a$ at late times ($0.1 \leq a \leq 1$) for TorC cosmology. The top panel shows the Planck only result, while the bottom panel shows the joint Planck + SH0ES fit. In both panels, the Planck 2018 best-fit \LCDM{} model is shown as a dashed line. Each curve is reconstructed from the corresponding posterior distribution using \texttt{fgivenx}~\cite{fgivenx}.}
\label{fig:hubble-evolution}
\end{figure}
\cref{fig:hubble-evolution} shows the posterior evolution of the Hubble parameter,~$H(a)$, for TorC cosmology between~$a = 0.1$ and~$a = 1$. In the Planck-only case (top panel), the Planck 2018~$\Lambda$CDM best-fit curve lies within the TorC posterior. A deviation is visible during the matter-dominated era, where TorC expands slightly more slowly than the~$\Lambda$CDM best-fit. When combining Planck 2018 with SH0ES (bottom panel), this feature becomes more pronounced: TorC remains lower than the~$\Lambda$CDM best-fit at early and intermediate times, but exceeds it at~$a = 1$, yielding a higher present-day Hubble parameter.
\hfill

\section{Prior compression}\label{sec:TorCprior}
\begin{figure}[h]
\centering
\includegraphics[width=\linewidth]{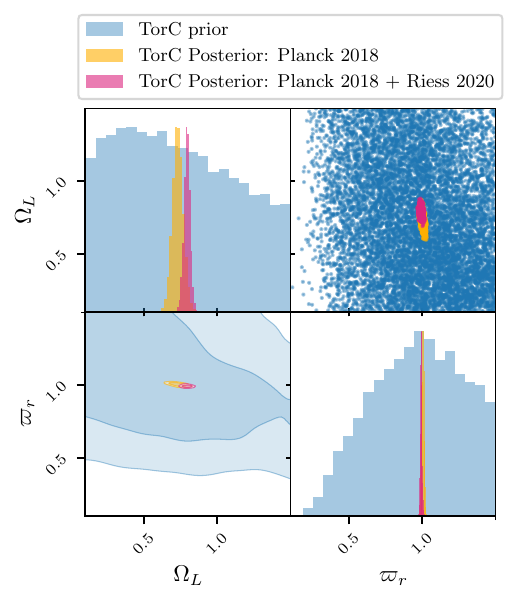}
\caption{A comparison of prior (blue) and posterior distributions for the TorC parameters~$\Omega_\Lambda$ and $\varpi_\mathrm{r}$, using the Planck 2018 data (yellow) and the joint Planck 2018 and SH0ES data (red). The prior distribution represents the additional parameter space introduced by TorC over the \LCDM{} model. The sharp compression into the posterior distribution illustrates the information gain, which is quantified by the KL divergence.}
\label{fig:TorCprior}
\end{figure}

Figure~\ref{fig:TorCprior} displays the prior compression for the TorC parameters $\Omega_\Lambda$ and $\varpi_\mathrm{r}$. The narrowing of the posterior distributions relative to the priors indicates that data strongly constrain these parameters. This information gain is quantitatively captured by the KL divergence values reported in~\cref{sec:model-comparison}.

\end{document}